\documentclass[a4paper,12pt, epsfig]{article}
\usepackage{epsfig}
\usepackage{amssymb,latexsym}
\usepackage{amsfonts}
\usepackage{amsmath}
\usepackage[colorlinks, linkcolor=blue, citecolor=blue, urlcolor=blue]{hyperref}
\usepackage{epstopdf}

\newskip\humongous \humongous=0pt plus 1000pt minus 1000pt

\newif\ifdtup

\catcode`\@=11

\@addtoreset{equation}{section}
\def\theequation{\thesection.\arabic{equation}}

\def\@normalsize{\@setsize\normalsize{15pt}\xiipt\@xiipt
\abovedisplayskip 14pt plus3pt minus3pt%
\belowdisplayskip \abovedisplayskip
\abovedisplayshortskip \z@ plus3pt%
\belowdisplayshortskip 7pt plus3.5pt minus0pt}

\def\small{\@setsize\small{13.6pt}\xipt\@xipt
\abovedisplayskip 13pt plus3pt minus3pt%
\belowdisplayskip \abovedisplayskip
\abovedisplayshortskip \z@ plus3pt%
\belowdisplayshortskip 7pt plus3.5pt minus0pt
\def\@listi{\parsep 4.5pt plus 2pt minus 1pt
      \itemsep \parsep
      \topsep 9pt plus 3pt minus 3pt}}

\relax

\catcode`@=12

\catcode`\@=11

\def\section{\@startsection{section}{1}{\z@}{3.5ex plus 1ex minus
    .2ex}{2.3ex plus .2ex}{\large\bf}}

\def\thesection{\arabic{section}}
\def\thesubsection{\arabic{section}.\arabic{subsection}}

\def\appendix{\setcounter{section}{0}
  \def\thesection{Appendix \Alph{section}}
  \def\thesubsection{\Alph{section}.\arabic{subsection}}
  \def\theequation{\Alph{section}.\arabic{equation}}}

\def\SymBoxes#1#2#3#4{\newdimen\un@t \un@t#3%
\raisebox{#1}{\rule{#2\un@t}{#4}\hskip-#2\un@t
\@tempdimb\un@t \advance\@tempdimb by-#4\@tempcntb#2\relax%
\@whilenum{\@tempcntb>0}\do{
\rule{#4}{\un@t}\hskip\@tempdimb \advance\@tempcntb by\m@ne}%
\hskip-#2\un@t \rule[\un@t]{#2\un@t}{#4}%
\rule[\un@t]{#4}{#4}\hskip-#4
\rule{#4}{\un@t}}\hskip-#4}                

\begin{document}


\newcommand{\dd}{\textrm{d}}

\newcommand{\beq}{\begin{equation}}
\newcommand{\eeq}{\end{equation}}
\newcommand{\bea}{\begin{eqnarray}}
\newcommand{\eea}{\end{eqnarray}}
\newcommand{\beas}{\begin{eqnarray*}}
\newcommand{\eeas}{\end{eqnarray*}}
\newcommand{\defi}{\stackrel{\rm def}{=}}
\newcommand{\non}{\nonumber}
\newcommand{\bquo}{\begin{quote}}
\newcommand{\enqu}{\end{quote}}
\newcommand{\tc}[1]{\textcolor{blue}{#1}}
\renewcommand{\(}{\begin{equation}}
\renewcommand{\)}{\end{equation}}
\def\de{\partial}
\def\Om{\ensuremath{\Omega}}
\def\Tr{ \hbox{\rm Tr}}
\def\rc{ \hbox{$r_{\rm c}$}}
\def\H{ \hbox{\rm H}}
\def\HE{ \hbox{$\rm H^{even}$}}
\def\HO{ \hbox{$\rm H^{odd}$}}
\def\HEO{ \hbox{$\rm H^{even/odd}$}}
\def\HOE{ \hbox{$\rm H^{odd/even}$}}
\def\HHO{ \hbox{$\rm H_H^{odd}$}}
\def\HHEO{ \hbox{$\rm H_H^{even/odd}$}}
\def\HHOE{ \hbox{$\rm H_H^{odd/even}$}}
\def\K{ \hbox{\rm K}}
\def\Im{ \hbox{\rm Im}}
\def\Ker{ \hbox{\rm Ker}}
\def\const{\hbox {\rm const.}}
\def\o{\over}
\def\im{\hbox{\rm Im}}
\def\re{\hbox{\rm Re}}
\def\bra{\langle}\def\ket{\rangle}
\def\Arg{\hbox {\rm Arg}}
\def\exo{\hbox {\rm exp}}
\def\diag{\hbox{\rm diag}}
\def\longvert{{\rule[-2mm]{0.1mm}{7mm}}\,}
\def\a{\alpha}
\def\b{\beta}
\def\e{\epsilon}
\def\l{\lambda}
\def\ol{{\overline{\lambda}}}
\def\ochi{{\overline{\chi}}}
\def\th{\theta}
\def\s{\sigma}
\def\oth{\overline{\theta}}
\def\ad{{\dot{\alpha}}}
\def\bd{{\dot{\beta}}}
\def\oD{\overline{D}}
\def\opsi{\overline{\psi}}
\def\dag{{}^{\dagger}}
\def\tq{{\widetilde q}}
\def\L{{\mathcal{L}}}
\def\p{{}^{\prime}}
\def\W{W}
\def\N{{\cal N}}
\def\hsp{,\hspace{.7cm}}
\def\hspp{,\hspace{.5cm}}
\def\hsppp{,\hspace{.25cm}}
\def\bo{\ensuremath{\hat{b}_1}}
\def\bfo{\ensuremath{\hat{b}_4}}
\def\co{\ensuremath{\hat{c}_1}}
\def\cfo{\ensuremath{\hat{c}_4}}
\def\th#1#2{\ensuremath{\theta_{#1#2}}}
\def\c#1#2{\hbox{\rm cos}(\th#1#2)}
\def\s#1#2{\hbox{\rm sin}(\th#1#2)}
\def\cp#1#2#3{\hbox{\rm cos}^#1(\th#2#3)}
\def\sp#1#2#3{\hbox{\rm sin}^#1(\th#2#3)}
\def\ctp#1#2#3{\hbox{\rm cot}^#1(\th#2#3)}
\def\cpp#1#2#3#4{\hbox{\rm cos}^#1(#2\th#3#4)}
\def\spp#1#2#3#4{\hbox{\rm sin}^#1(#2\th#3#4)}
\def\t#1#2{\hbox{\rm tan}(\th#1#2)}
\def\tp#1#2#3{\hbox{\rm tan}^#1(\th#2#3)}
\def\m#1#2{\ensuremath{\Delta M_{#1#2}^2}}
\def\mn#1#2{\ensuremath{|\Delta M_{#1#2}^2}|}
\def\u#1#2{\ensuremath{{}^{2#1#2}\mathrm{U}}}
\def\pu#1#2{\ensuremath{{}^{2#1#2}\mathrm{Pu}}}
\def\an{\ensuremath{\alpha_n}}
\newcommand{\C}{\ensuremath{\mathbb C}}
\newcommand{\Z}{\ensuremath{\mathbb Z}}
\newcommand{\R}{\ensuremath{\mathbb R}}
\newcommand{\rp}{\ensuremath{\mathbb {RP}}}
\newcommand{\vac}{\ensuremath{|0\rangle}}
\newcommand{\vact}{\ensuremath{|00\rangle}                    }
\newcommand{\oc}{\ensuremath{\overline{c}}}
\renewcommand{\cos}{\textrm{cos}}
\renewcommand{\sin}{\textrm{sin}}
\renewcommand{\cot}{\textrm{cot}}

\newcommand{\Vol}{\textrm{Vol}}

\newcommand{\half}{\frac{1}{2}}

\def\changed#1{{\bf #1}}

\begin{titlepage}

\def\thefootnote{\fnsymbol{footnote}}

\begin{center}
{\large {\bf
The Neutrino Mass Hierarchy at Reactor Experiments\\now that $\theta_{13}$ is Large
  } }

\bigskip

\bigskip

{\large \noindent Emilio
Ciuffoli$^{1}$\footnote{ciuffoli@ihep.ac.cn}, Jarah
Evslin$^{1}$\footnote{\texttt{jarah@ihep.ac.cn}} and Xinmin Zhang$^{2,
1}$\footnote{\texttt{xmzhang@ihep.ac.cn}} }
\end{center}

\renewcommand{\thefootnote}{\arabic{footnote}}

\vskip.7cm

\begin{center}
\vspace{0em} {\em  { 1) TPCSF, IHEP, Chinese Acad. of Sciences\\
2) Theoretical physics division, IHEP, Chinese Acad. of Sciences\\
YuQuan Lu 19(B), Beijing 100049, China}}

\vskip .4cm

\vskip .4cm

\end{center}

\vspace{1.0cm}

\noindent
\begin{center} {\bf Abstract} \end{center}

\noindent
Now that $\theta_{13}$ is known to be large, a medium baseline reactor experiment can observe the fine structure of the electron antineutrino survival probability curve, approximately periodic oscillations in $L/E$ with wavelength $4\pi/\mn31$.  The periodicity with respect to $L/E$ is broken by 2-3 oscillations which, in the case of the normal (inverted) hierarchy, shift the first 16 oscillations nearly 1\% higher (lower) and move the next 16  lower (higher).  The energy of each peak determines a particular combination of the mass differences, for example $\cp212\mn31+\sp212\mn32$ for all peaks visible at baselines under 40 km.  Comparing these combinations with each other or with NO$\nu$A results one can in principle determine the mass hierarchy.  Alternately, as the Fourier transforms of the 1-3 and 2-3 oscillation probabilities are out of phase by the 1-2 oscillation probability, near the maximum of the 1-2 oscillation the complex phase of the total survival probability can be used to determine the hierarchy.  Two interference effects make this task difficult.  First, kilometer distances between the reactors reduce the amplitudes of peaks below about 4 MeV.   Second, even reactors 100 or more kilometers away significantly obscure the 1-2 oscillation maximum, which also complicates a measurement of the solar mixing angle with a single detector.

\vfill

\begin{flushleft}
{\today}
\end{flushleft}
\end{titlepage}

\hfill{}


\setcounter{footnote}{0}

\section{Introduction}

10 years ago Petcov and Piai, on the basis of the existing experimental indications that the solar mass splitting $\Delta M^2_{21}\sim 2.4 \times10^{-4} \rm{eV}^2$, suggested that a 20-25 km baseline neutrino detector may be able to determine the neutrino mass hierarchy from reactor antineutrinos if the solar mass splitting $\Delta M^2_{21}$ is greater than about $10^{-4} \rm{eV}^2$ \cite{petcovidea}.  It is now known that $\Delta M^2_{21}\sim 7.5\times 10^{-5} \rm{eV}^2$ \cite{gando}, implying that  the difference between the normal and inverted hierarchies cannot be seen at a baseline below about 40 km.  The low fluxes at these long baselines led various groups \cite{hawaii,caojun} to conclude that the individual peaks in the neutrino spectrum would be difficult to resolve, and so the only hope for a hierarchy determination would be to sum them via a Fourier transform.  Even in this case the necessary detectors were extremely large and the experiments slow.

This all changed with the recent determination of $\theta_{13}$ \cite{daya,reno} which, as was hinted last year \cite{doublechooz,globale}, corresponds to a value of $\spp2213$ up to 10 times higher than had been considered by the authors of Refs.~\cite{hawaii,caojun}.  The larger mixing angle increases the sizes of the oscillations used to determine the hierarchy by an order of magnitude.  This means both that a reactor experiment to determine the hierarchy is now practical and also that the analysis of the optimal baseline and experimental configuration must now be redone.  Indeed, such a reanalysis is urgent as such an experiment will be built soon \cite{caojunseminario,renonuturn,yifangseminario}.

We will perform this updated analysis in three papers, mirroring the structure of Ref.~\cite{caojun}.   In this first paper we will analyze a medium baseline reactor experiment analytically.  We will derive old observations relating observables of the Fourier transformed spectrum to the neutrino mass hierarchy and we will also provide new methods for determining various combinations of the neutrino masses and the hierarchy.  The optimal analysis will be a combination of the old and the new to be determined by simulations.  We will also discuss problems which have so far escaped attention in the literature.   In the second paper \cite{noispurioso} we discuss the spurious dependence, first observed in Ref.~\cite{oggi}, of the parameters of Ref.~\cite{caojun} on the theoretical reactor spectrum and on the mass difference $\mn32$ and we show that it can be removed with a weighted Fourier transform.  Finally in the third paper \cite{noisim} will describe the results of our simulations, indicating for example the optimal baseline and the optimal location for such an experiment in China's Guangdong province.

A central theme in these papers will be the observation that, since reactors are typically separated by of order 1 km in a reactor array, the baselines of neutrinos from different reactors are different.  As a result the oscillations of low energy neutrinos, which perform half an oscillation while traveling from one reactor to the next, will be invisible at the detector as the neutrinos from one reactor arrive at the oscillation maximum while the other is at its minimum.  We will show that, due to a degeneracy between the hierarchy and a mass difference, a determination of the neutrino mass hierarchy is impossible in an experiment which cannot resolve these low energy peaks.  Thus the angle between the detector and the lines extending between reactors is an essential variable in the determination of the optimal detector location.  For example, for a linear array of reactors like RENO or Daya Bay plus Ling Ao, this effect can be eliminated if the detector is placed orthogonal to the array.

The greatest sensitivity to the hierarchy arises near the 1-2 oscillation maximum, at about 60 km.  However the flux from a distant reactor at the 1-2 oscillation minimum of 120 km will dominate over the flux from the near reactor in this region.  In fact, this will be the case for a detector placed 60 km away from Daya Bay and Ling Ao in the orthogonal direction, as the proposed Haifeng and Huizhou reactors will be near the 1-2 minimum.  Also if a detector is placed equidistant from Daya Bay and Haifeng or Huizhou at the position suggested in Ref. \cite{caojunseminario} or \cite{yifangseminario} then the 1-2 maximum neutrinos from Daya Bay and Haifeng or Huizhou will correspond to the 1-2 minimum for neutrinos from the proposed reactor at Lufeng \cite{yifangseminario}.  This not only makes the determination of the hierarchy more difficult, but also is detrimental to the sensitivity to $\theta_{12}$.  The ideal solution to this problem is to use two detectors at different distances, say 40 and 70 km.  However a more economical solution is to keep the baselines short so that the flux from the desired reactor complex dominates over the fluxes from others, for example one can consider a baseline of 45-50 km. 

We will begin in Sec. \ref{teorsez} with a review of standard results on 3-flavor neutrino oscillations and the electron antineutrino survival probability.  We describe the interference between the 1-3 and 2-3 oscillations \cite{petcovidea} which leads to beats at the 1-2 frequency and we numerically find the energies of the combined peaks.  Then in Sec. \ref{masssez} we describe how the positions of various peaks can be used to obtain combinations of the neutrino mass differences.  The position of the $n$th peak determines its own combination $\Delta M^2_{(n)}$.   We will see that, as the first 10 peaks all determine the same mass combination, they do not allow a determination of the mass hierarchy, but the next 5 do, which is why short baselines are not sufficient for a determination of the hierarchy.  In Sec. \ref{vincsez} we discuss the consequences of the finite energy resolution and the finite neutrino flux.  Both provide obstacles to locating the $n>10$ peaks, and so to a determination the hierarchy.  In Sec. \ref{fouriersez} we discuss analyses of the Fourier transform of the neutrino spectrum.  These have the advantage that, when the nonlinearity of the detector response is well understood, they sum the peaks together, and so render the signal stronger. We rederive three old ways in which the hierarchy can be determined from this transformed spectrum and add two new methods to the list. 

Finally in Sec. \ref{intsez} we discuss the consequences of the fact that not all of the neutrinos detected traveled the same distance.  The distances may differ by of order a kilometer because the individual reactors in an array are not coincident, leading to an interference effect which greatly diminishes the amplitudes of the low energy, high $n$, peaks. We will see that this interference can be avoided if the detector is placed perpendicular to the array.   Also, while it has long been known \cite{hawaii} that neutrinos from reactors at the 1-2 oscillation maximum baseline are the most useful for determining the hierarchy, we will see that this signal can be overwhelmed by neutrinos from distant reactors at the 1-2 minimum, and we will argue that this is indeed the case if a detector is placed at the 1-2 maximum orthogonal to the Daya Bay, Ling Ao reactor array.  This problem, as well as the related error in a determination of $\theta_{12}$, can be reduced by shortening the baseline or, if possible, adding another detector at a different baseline.

Today we received a preprint \cite{oggi} which has significant overlap with the results in our Sec. \ref{teorsez}.  For example their $\phi$, introduced in Ref.~\cite{parke2007}, is proportional to our $\an/n$.

\section{The electron neutrino survival probability} \label{teorsez}

\subsection{Short and long oscillations}

The electron neutrino weak interaction eigenstate $|\nu_e\rangle$ is not an energy eigenstate $|k\rangle$, but it can be decomposed into a real sum of energy eigenstates
\beq
|\nu_e\rangle=\c12\c23|1\rangle+\s12\c13|2\rangle+\s13|3\rangle.
\eeq
In the relativistic limit, after traveling a distance $L$, the survival probability of a coherent electron (anti)neutrino wavepacket with energy $E$ can be expressed in terms of the mass matrix $\mathbf{M}$
\bea
P_{ee}&=&|\langle\nu_e|\mathrm{exp}\left(i\frac{\mathbf{M}^2L}{2E}\right)|\nu_e\rangle|^2\label{pee}\\
&=&\sp413+\cp412\cp413+\sp412\cp413+\frac{1}{2}(P_{12}+P_{13}+P_ {23})\nonumber\\
P_{12}&=&\spp2212\cp413\cos\left(\frac{\m21L}{2E}\right)\hsp
P_{13}=\cp212\spp2213\cos\left(\frac{\mn31L}{2E}\right)\nonumber\\
P_{23}&=&\sp212\spp2213\cos\left(\frac{\mn32L}{2E}\right)\nonumber
\eea
where $\m{i}{j}$ is the mass squared difference of mass eigenstates $i$ and $j$.  Notice that the survival probability is a sum of cosines and so its cosine Fourier transform with respect to the variable $L/E$ is just a sum of delta functions whereas its sine transform vanishes.

The three cosines in the survival probability (\ref{pee}) identify two characteristic frequencies of the $L/E$ oscillations.  The $P_{12}$ term oscillates at a low frequency 
\beq
\frac{\m21}{2}\sim 3.8\times 10^{-5}\ \mathrm{eV}^2\sim 0.19\ \mathrm{MeV/km} .
\eeq
Therefore the maximum 1-2 oscillation occurs at
\beq
\frac{L}{E}=\frac{\pi}{\m21/2}\sim 17 \ \mathrm{km/MeV}. \label{unodue}
\eeq
A medium baseline reactor experiment, with a baseline of under 100 km,  can observe at most one or two such oscillations.  Instead such experiments will focus on shorter oscillations, characterized by the $P_{13}$ term, which have frequency
\beq
\frac{\mn31}{2}\sim 1.2\times 10^{-3}\ \mathrm{eV}^2\sim 6.1\ \mathrm{MeV/km} 
\eeq
corresponding to a wavelength of
\beq
\Delta\left(\frac{L}{E}\right)=\frac{2\pi}{\mn31/2}\sim 1.03 \ \mathrm{km/MeV}. \label{corto}
\eeq
At a medium baseline reactor one may hope to see 5 to 15 such oscillations.

What about the $P_{23}$ term  in (\ref{pee})?  The frequency is $\mn32/2$, which is about 3\% more or less than $\mn31/2$ depending on the hierarchy.  However the amplitude is less than that of $P_{13}$ by a factor of
\beq
a=\tp212\sim 0.45.
\eeq
As the frequencies of the two short oscillations are similar but $P_{23}$ has a smaller amplitude, the total short distance oscillation $P_{13}+P_{23}$ is a deformation of $P_{13}$ alone.  However the 2-3 oscillations serve to slightly displace the 1-3 peaks and shift the amplitudes with a pattern which repeats at the beat frequency $\m21/2$.

More precisely, while the $n$th maximum of $P_{13}$ is at $L/E=4\pi n/\mn31$, the $n$th peak of $P_{13}+P_{23}$ is at
\beq
\frac{L}{E}=\frac{4\pi}{\mn31}(n\pm\an) \label{pichi}
\eeq
for a vector $\an$ which is determined entirely by the neutrino mass matrix.  The positive (negative) sign applies to the normal (inverted) hierarchy.  As the derivatives of the reactor neutrino spectrum and the 1-2 oscillations are small, the peaks of the total survival probability $P_{ee}$ and its product with the reactor neutrino spectrum, which is the observed neutrino spectrum, are roughly located at the values given in Eq. (\ref{pichi}).

From Eq. (\ref{pichi}) it is possible to understand the main obstruction to determinations of the hierarchy at short baselines.  The position of the $n$th peak is only sensitive to the mass combination
\beq
\Delta M^2_{eff}=\frac{\mn31}{1\pm\an/n}.
\eeq
As we will see (analytically and then numerically in Fig. \ref{anfig}) at low $n$, corresponding to peaks visible at relatively short baselines, the values of $\an$ are roughly linear.  Thus in this regime the positions of all of the peaks are sensitive to the same effective mass and so are independent of the hierarchy so long as that effective mass applies: A change in the hierarchy can be compensated by a shift in $\mn31$ while leaving the peaks fixed. {\textbf{The hierarchy can only be determined from the nonlinearity of $\an$.}}

\subsection{Finding $\an$}

The vector $\an$ encodes the effect of the neutrino mass matrix on the location of the survival probability peaks.  Therefore a knowledge of this dependence together with a measurement of the peaks allows one to reconstruct some elements of the mass matrix.

The values of $\an$ are determined from (\ref{pichi}) by first finding the extrema of $P_{13}+P_{23}$
\bea
0&=&\frac{\partial}{\partial E}(P_{13}+P_{23})\propto \frac{\partial}{\partial E}\left[\cos\left(\frac{\mn31L}{2E}\right)+\tp212\cos\left(\frac{\mn32L}{2E}\right)\right]\\
&\propto&\sin\left(\pm 2\pi\an\right)+\left(1\mp\epsilon\right)\tp212\sin\left(2\pi[\pm\an
\mp\epsilon(n\pm\an)]\right)\hsp
\epsilon=\frac{\m21}{\mn31}\nonumber
\eea
where the upper sign applies to the normal hierarchy.

As $\epsilon<<1$ and $\an<<n$ we may approximate
\beq
0\sim \sin(2\pi\an)+\tp212\sin(2\pi[\an-\epsilon n]). \label{alfeq}
\eeq
This can be expanded in a power series in $n$.  The highest energy peaks occur at small values of $n$, where the linear term in this expansion suffices
\beq
0\sim 2\pi(1+\tp212)\an- 2\pi\tp212\epsilon n
\eeq
which is easily solved for $\an$
\beq
\an\sim\epsilon\sp212 n\sim  0.010 n.
\eeq

Thus the n$th$ peak, for $n$ sufficiently small, lies at
\beq
\frac{L}{E}\sim\frac{4\pi}{\mn31}(1\pm\epsilon\sp212) n\sim \frac{4\pi n}{\mn31\mp\sp212\m21} \label{degen}
\eeq
where the upper sign corresponds to the normal hierarchy.  

The basic problem facing shorter baseline experiments, which are only sensitive to peaks at small $n$, is already apparent in Eq. (\ref{degen}).  The mass difference $\mn31$ is degenerate with the hierarchy
\beq
\mn31(\mathrm{normal})= \mn31(\mathrm{inverted})+2\sp212\m21 .
\eeq
Therefore any experiment with such a short baseline that all observable peaks have $n$ in the regime in which $\an$ is linear is, alone, incapable of determining the hierarchy.  Such an experiment can, however, determine the combination 
\beq
\Delta M^2_{eff}=\mn31\mp\sp212\m21=\cp212\mn31+\sp212\mn32. \label{meff}
\eeq
This is just the short baseline effective mass first reported in Ref.~\cite{parke2005}.

Therefore reactor experiments can determine the hierarchy in only two ways.  Either an accurate determination of $\Delta M^2_{eff}$ can be combined with an accurate determination of another combination of the mass differences obtained from another experiment, a difference which needs to be known more precisely than the atmospheric mass difference measured by MINOS, or else the experiment needs to be sensitive to peaks at large enough $n$ that the linear approximation breaks down.  So just how large does $n$ need to be?

\subsection{Cubic terms is $\an$} \label{cubicosez}
To see where interference between 2-3 and 1-3 oscillations pushes the peaks of $P_{13}+P_{23}$ at larger values of $n$, we need to expand \an\ to cubic order
\beq
\an\sim  \epsilon\sp212 n+b n^3 \label{cubico}
\eeq
and to substitute this expansion into Eq. (\ref{alfeq}).   The linear term in $\an$ already solves this equation at linear order.  At cubic order it yields
\beq
0\sim 2\pi(1+\tp212)b -\frac{4\pi^3}{3}\epsilon^3\sp212(\sp212-\cp212)       
\eeq
and so
\beq
b\sim\frac{2\pi^2}{3}\epsilon^3\sp212\cp212(\sp212-\cp212) \sim - 1.6\times 10^{-5} .
\eeq

The linear approximation to $\an$ is reliable when the cubic term in Eq. (\ref{cubico}) is much smaller than the linear term
\beq
n<<\sqrt{\frac{\epsilon\sp212}{|b|}}\sim 25.
\eeq
For example, at the 10th peak the contribution of the cubic term to the energy is only one sixth of that of the linear term.  The linear term we have seen shifts the effective mass by about 1\%, thus the cubic term, which is the leading hierarchy-dependent term, only shifts the energy of the tenth peak by about one sixth of a percent.  The other hierarchy would lead to a shift of a sixth percent in the other direction, so overall the difference in the energies of the 10th peaks in the normal and inverted hierarchy is about one third of a percent.

Therefore if only the first ten peaks can be measured at a given baseline, the detector will need to be able to determine the position of the tenth peak with a precision of a half percent for only a one sigma determination of the hierarchy, making a determination of the hierarchy using such an experiment alone quite unlikely.

\begin{figure} 
\begin{center}
\includegraphics[width=5in,height=2in]{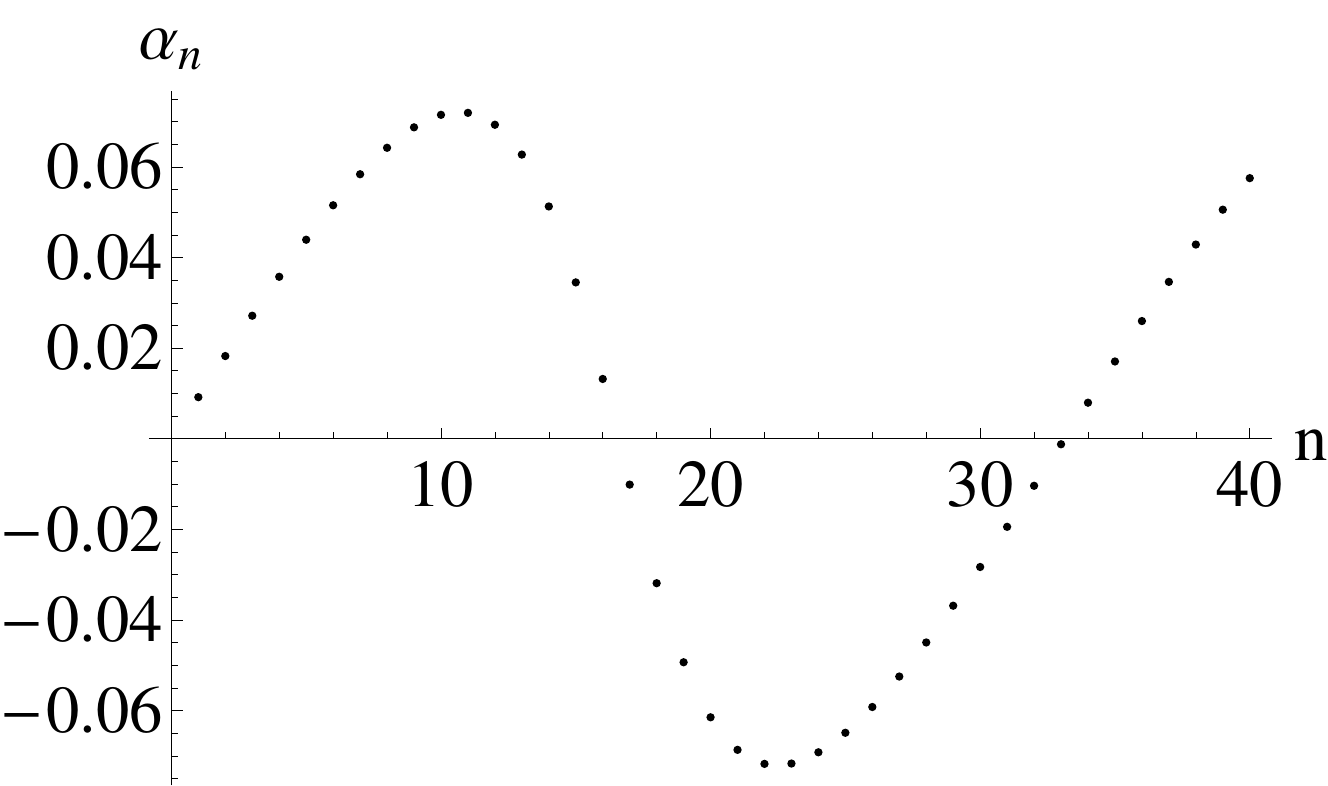}
\caption{$\an$ determined numerically}
\label{anfig}
\end{center}
\end{figure}

The cubic expansion is no longer reliable for higher peaks, but $\an$ can be determined numerically.  As seen in Fig. \ref{anfig}, it is periodic modulo $1/\epsilon$ and is zero at every multiple of $1/2\epsilon$.  But the main problem is that it is nearly linear for $n<10$, which is why the hierarchy is so nearly degenerate with the shift in the mass differences at all of these peaks.

A $2/k$ percent precision measurement of the peak energy of the 14th peak would also give a $k$ sigma indication of the hierarchy.  The 14th peak can only be seen if it occurs at a high enough energy that the neutrino flux and detector resolution are sufficient to discern it.  For example if one requires it to appear above 3 MeV, so that the detector resolution may be better than 2\%, then using Eq. (\ref{corto}) the minimum baseline is about
\beq
L_{\mathrm{min}}\sim 45\ \mathrm{km}.
\eeq

\section{Determining mass differences from peak positions} \label{masssez}

\subsection{The reactor neutrino energy spectrum}

In the previous section we saw that the aperiodicity of the peaks in the electron (anti)neutrino survival probability is determined by the neutrino mass spectrum.  In particular, in $L/E$ space the peak positions are periodic modulo $1/\epsilon\sim 32$ and each set of $1/2\epsilon$ peaks is displaced about 1 percent either left or right depending on the hierarchy.  Thus, to determine the hierarchy, it suffices to measure the positions of enough peaks to within 2 percent precision.

The trouble with such a procedure is that no single experiment has access to all of the peaks, as there is a single useful baseline per detector and the energy spectrum is limited to that which is produced by a nuclear reactor, which leads to a maximum usable neutrino energy.  Not even all of these energies are accessible as each type of neutrino detector has a minimum energy which it is able to detect.   To determine which peaks may be seen by a particular experiment, one must combine both of these constraints.

The neutrino flux from a reactor results almost entirely from decays of just 4 isotopes: \u35,\ \pu39,\ \pu41\ and \u38.  The flux $\phi_i(E)$ of neutrinos from each isotope $i$ is traditionally approximated as the exponential of a polynomial in the neutrino energy $E$ \cite{vogelengel}
\beq
\phi_i(E)=\mathrm{exp}\left(\sum_{k=1}^{m} a_{ki}E^{k-1}\right). \label{polinom}
\eeq
Theoretical errors on these fluxes are often claimed to be near the 2-3\% level, although recent theoretical fluxes \cite{nuovoflusso} appear to be about 6\% higher than the fluxes measured at very short baseline experiments \cite{reattoreanom} and at 1 kilometer experiments \cite{noiunokm}.  

The precision of such a phenomenological law depends on the degree $n$ of the fit polynomial.  For neutrinos with between 2 and 7.5 MeV, which will be the ones of interest in this note, it was shown in Ref. \cite{huber2004} that a quadratic ($m=3$) fit tends to introduce errors of order 2-3\% whereas a 6 parameter ($m=6$) fit introduces errors of order 1\%, well below the error in the theoretical flux.  The differences between these parameterizations are oscillations over a characteristic scale of 1-2 MeV, which are likely too broad to give a false signal for a 1-3 oscillation peak, but may well disguise the depth of 1-2 oscillations and so affect the measured value of $\theta_{12}$.  

The most recent theoretical estimate of the flux is in Ref. \cite{huberflusso}, which shows a systematic 3\% excess over the previous estimates \cite{nuovoflusso} at energies above 6 MeV and a 1\% surplus beyond 4 MeV, which is within the theoretical errors of the calculations.  This correction again will affect a single detector determination of $\theta_{12}$.

Not all of the neutrinos which are generated by a reactor will be measured.   The maximum number of neutrinos which can be measured at a given energy $E$ is the product of the produced flux with the fraction of neutrinos at that energy which can be measured.  For example, if neutrinos are measured via the inverse $\beta$ decay reaction
\beq
\overline{\nu}_e+p\rightarrow n+e^+
\eeq
then the maximum number of neutrinos detected is the flux/area at the baseline $L$ multiplied by the inverse $\beta$ decay cross section which at tree level is \cite{sezionedurto}
\beq
\sigma(E)=0.0952\times 10^{-42}\mathrm{cm}^2(E_e p_e/\mathrm{MeV}^2) \label{albero}
\eeq
where the positron energy and momenta are, ignoring the neutron recoil, given in terms of the neutron, proton and electron rest masses
\beq
E_e=E-m_n+m_p+m_e\sim E-780\ \mathrm{keV}\hsp
p_e=\sqrt{E_e^2-m_e^2}. \label{posenergia}
\eeq

\begin{figure} 
\begin{center}
\includegraphics[width=5in,height=2in]{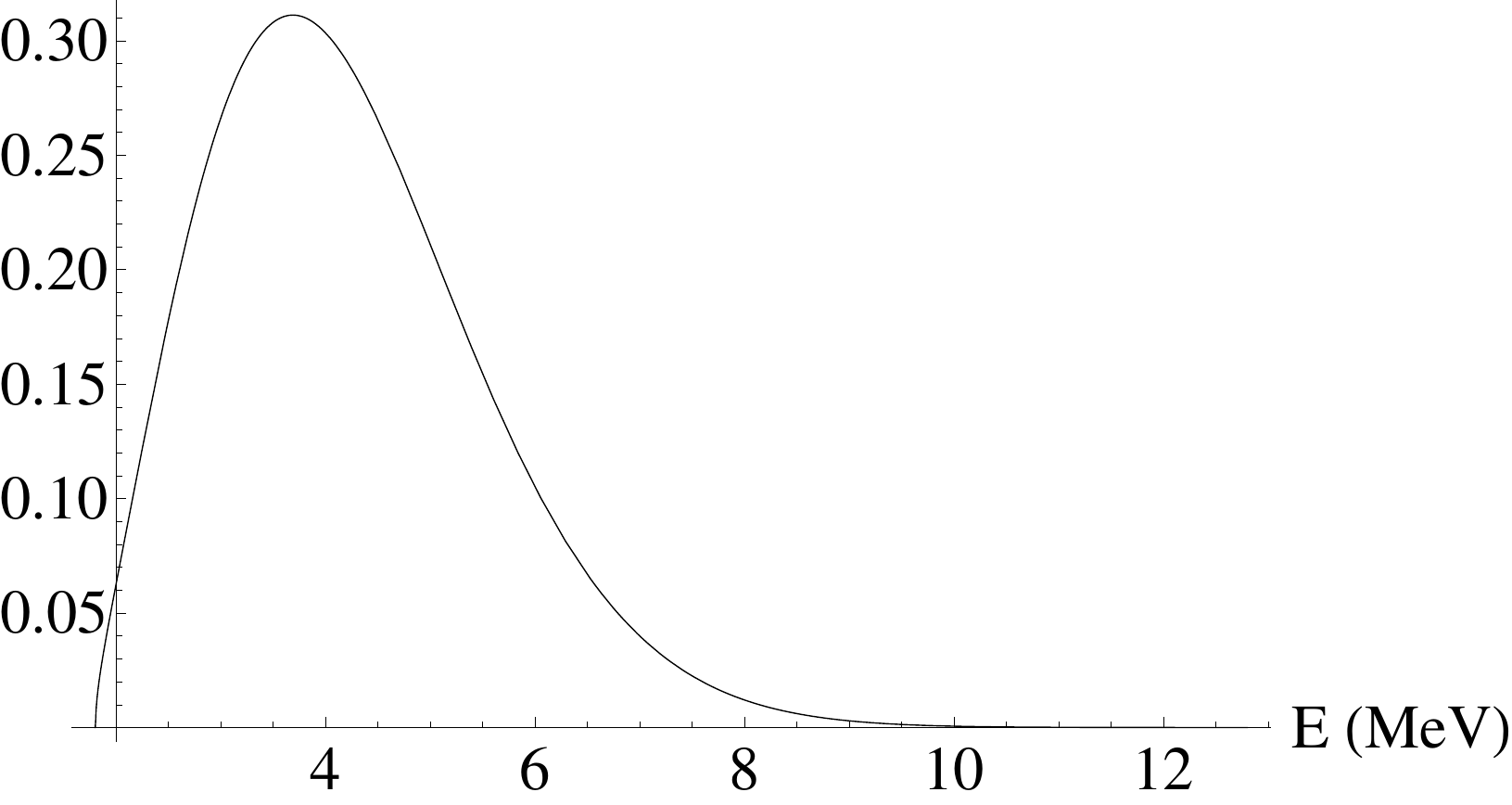}
\caption{The theoretical reactor neutrino spectrum as measured with inverse $\beta$ decay.}
\label{flussoorig}
\end{center}
\end{figure}

Combining the reactor flux (\ref{polinom}), with the coefficients $a_{ki}$ and typical isotope fractions of Ref. \cite{vogelengel}, with the tree level cross section (\ref{albero}), one obtains the theoretical reactor flux depicted in Fig. \ref{flussoorig}.  While inverse $\beta$ decay is kinematically forbidden if $E<m_n+m_e-m_p\sim 1.8$\ MeV, it can be seen in Fig. \ref{flussoorig} that the flux/energy is maximized at 3.6 MeV, falls to one third of its maximum by 6 MeV and to 10\% of its maximum by 7.5 MeV.  Thus useful information can only be obtained about the spectrum for energies within a factor of 2, which for each detector corresponds to values of $L/E$ and so $n$ within about a factor of 2.

\subsection{Effective masses at various baselines}

When neutrino oscillations are included, the theoretical flux $\Phi(E)$ from a reactor at a distance $L$ is then multiplied by the survival probability $P_{ee}$ given in Eq. (\ref{pee})
\beq
\Phi(E)=\sum_i c_i \phi_i(E)\sigma(E) P_{ee}(L/E)
\eeq
where $c_i$ is the quantity of each isotope in the reactor.

\begin{figure} 
\begin{center}
\includegraphics[width=3.2in,height=2in]{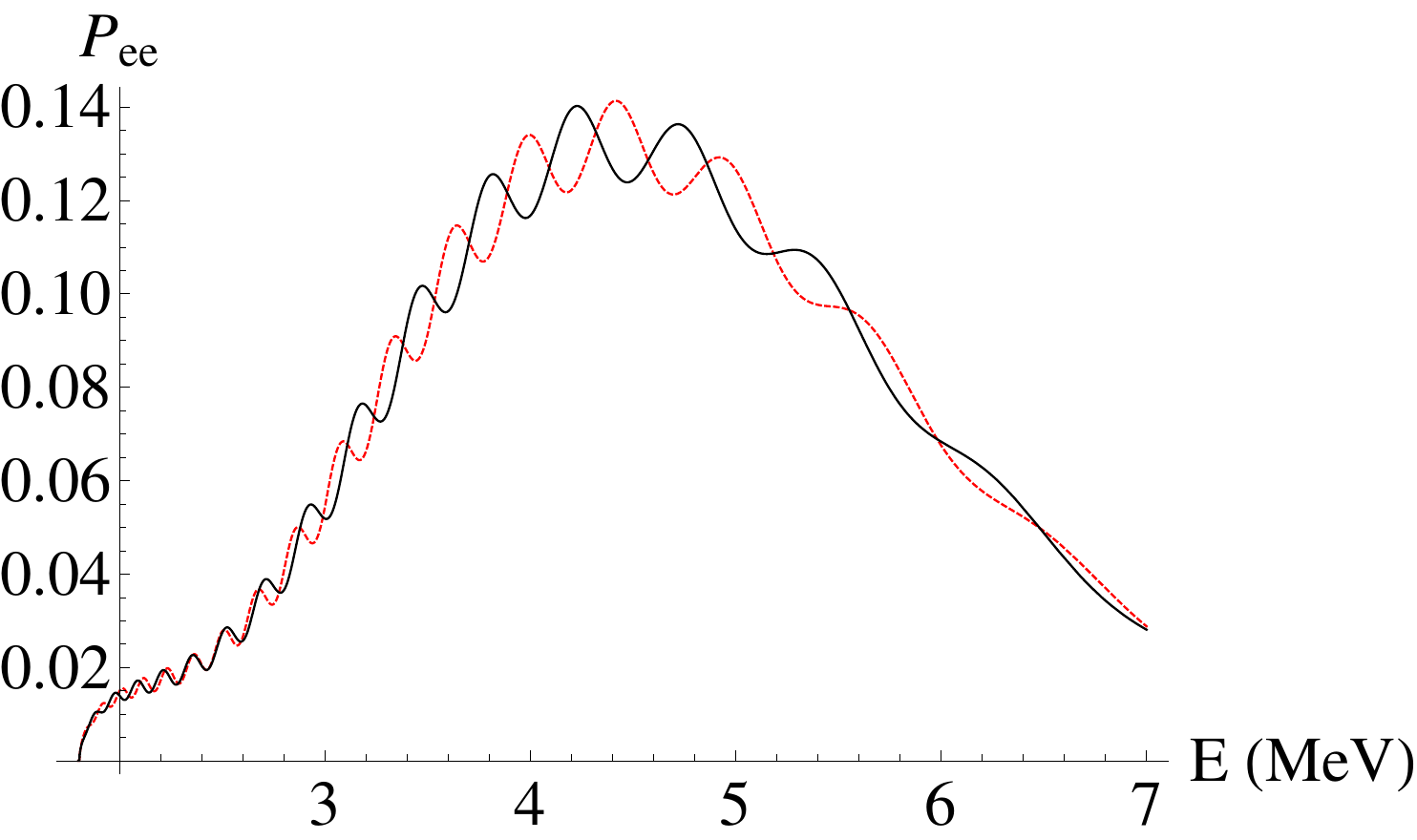}
\includegraphics[width=3.2in,height=2in]{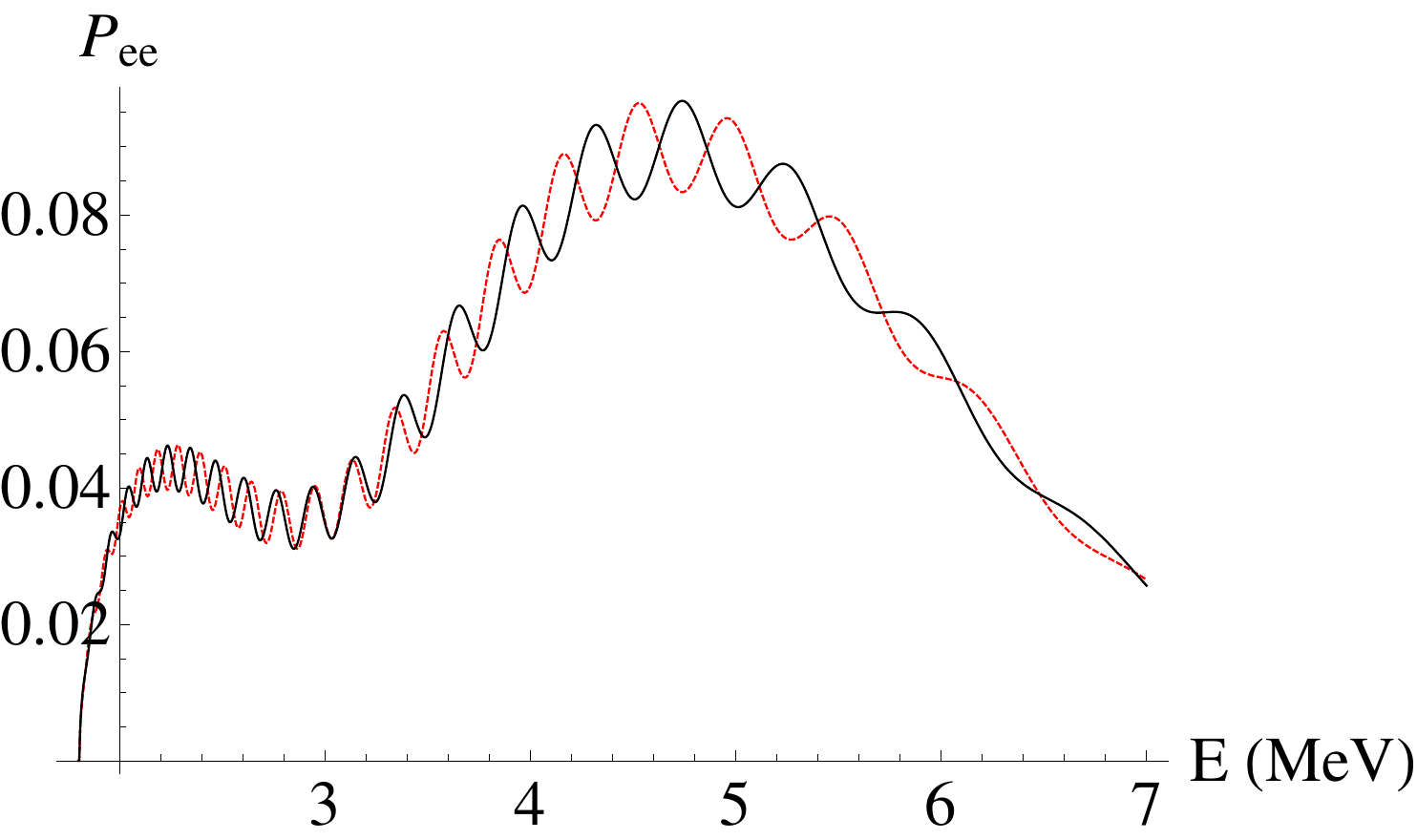}
\includegraphics[width=3.2in,height=2in]{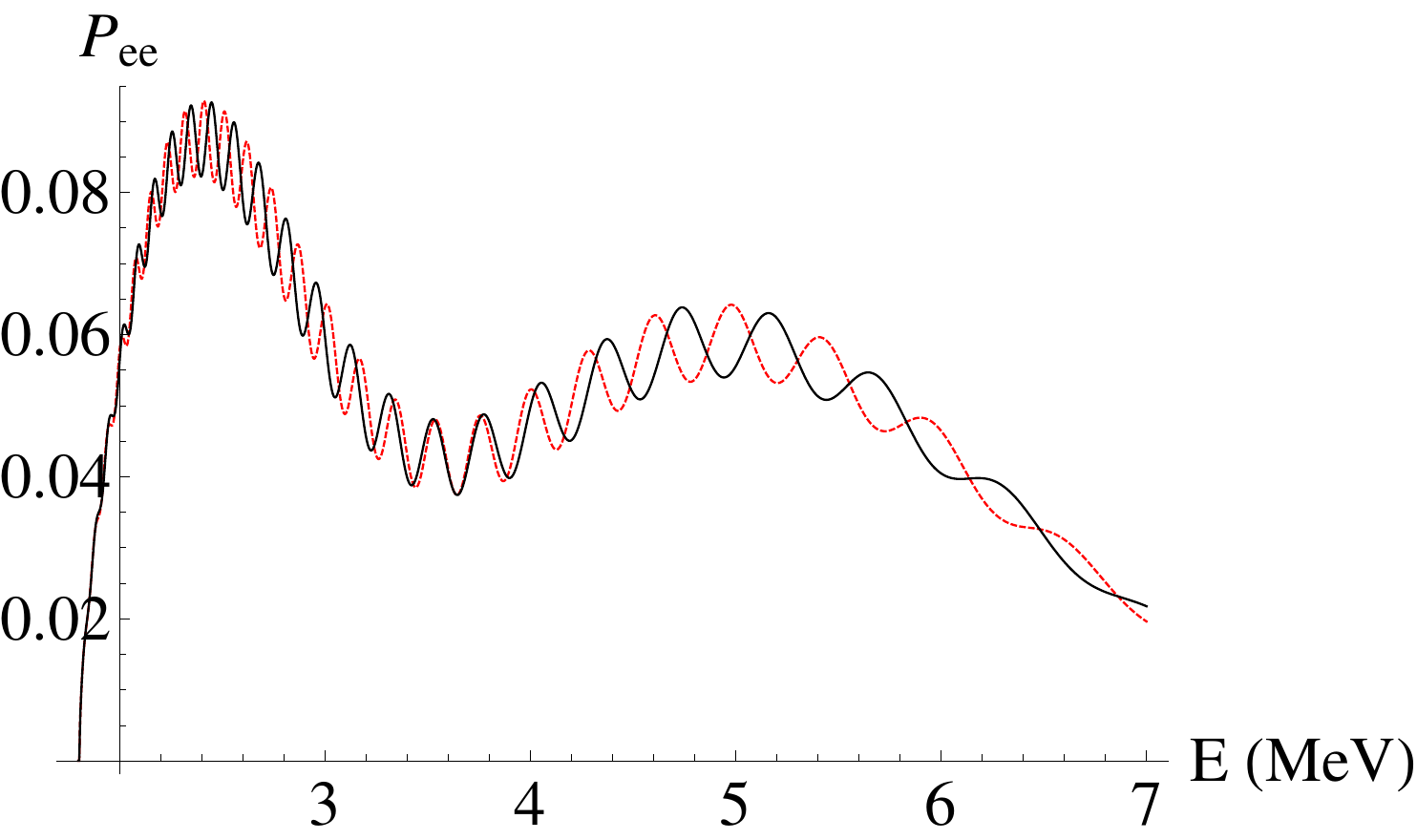}
\includegraphics[width=3.2in,height=2in]{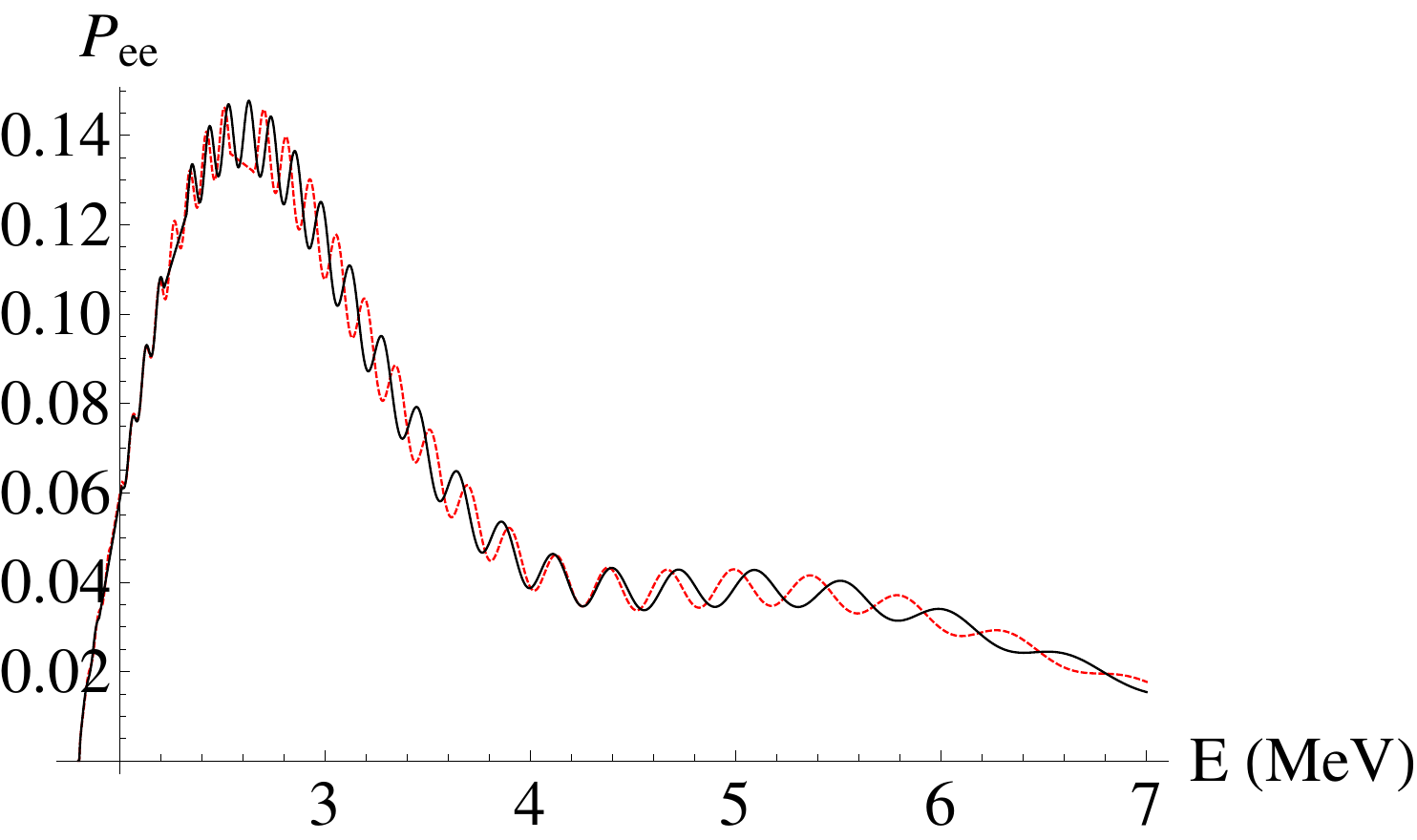}
\caption{Theoretical neutrino fluxes, including 3 flavor oscillation, for the normal (solid curve) and inverted (dashed curve) hierarchies as seen at 40, 50, 60 and 70 km.}
\label{tuttiflussi}
\end{center}
\end{figure}

We will use $\m21$ and $\spp2212$ from Ref. \cite{gando}, $\mn32$ determined by combining neutrino and antineutrino mass differences from Ref. \cite{minosneut2012}  and $\spp2213$ from \cite{neut2012} 
\beq
\m21=7.50\times 10^{-5}{\mathrm{\ eV^2}}\hsppp
\mn32=2.41\times 10^{-3}{\mathrm{\ eV^2}}\hsppp
\spp2212=0.857\hsppp
\spp2213=0.089.
\eeq
$\mn31$ is determined using the normal and inverted hierarchies.  The resulting neutrino flux is shown in Fig. \ref{tuttiflussi} at baselines of 40, 50, 60 and 70 km.

The numbers $n$ of the local maxima can be read from (\ref{pichi}) by approximating $\mn31\sim\mn32$ and setting $\an=0$
\beq
n\sim\frac{\mn32}{4\pi}\frac{L}{E}\sim 0.97 \frac{L/\textrm{km}}{E/\textrm{MeV}}. \label{nvalore}
\eeq
As the error on $\mn32$ is of order 3\% and the difference between $\Delta M^2_{eff}$ and $\mn32$ is also of order 2\%, one may expect an error of 3-5\% in Eq. (\ref{nvalore}).  The fractional energy difference between the $n$th and $(n+1)$st peak is $1/n$.  Therefore an optimal detector can robustly determine $n$ given a single peak only if $n$ is less than about 10.  Limitations placed by the finite energy resolution of the detector are discussed in Subsec. \ref{ressez}.

Although the energies of peaks in Fig. \ref{tuttiflussi} are strongly dependent upon the hierarchy at every baseline, this does not mean that a measurement of the spectra can actually allow one to determine the hierarchy.  The problem, as was discovered in Ref.~\cite{parke2007} in this context and as we will describe in Sec. \ref{teorsez}, is that the energies of the first 10 or so peaks only determine the mass difference $\Delta M^2_{eff}$ of Eq (\ref{meff}), leaving the hierarchy degenerate with $\mn31$.  

\begin{figure} 
\begin{center}
\includegraphics[width=3.2in,height=2in]{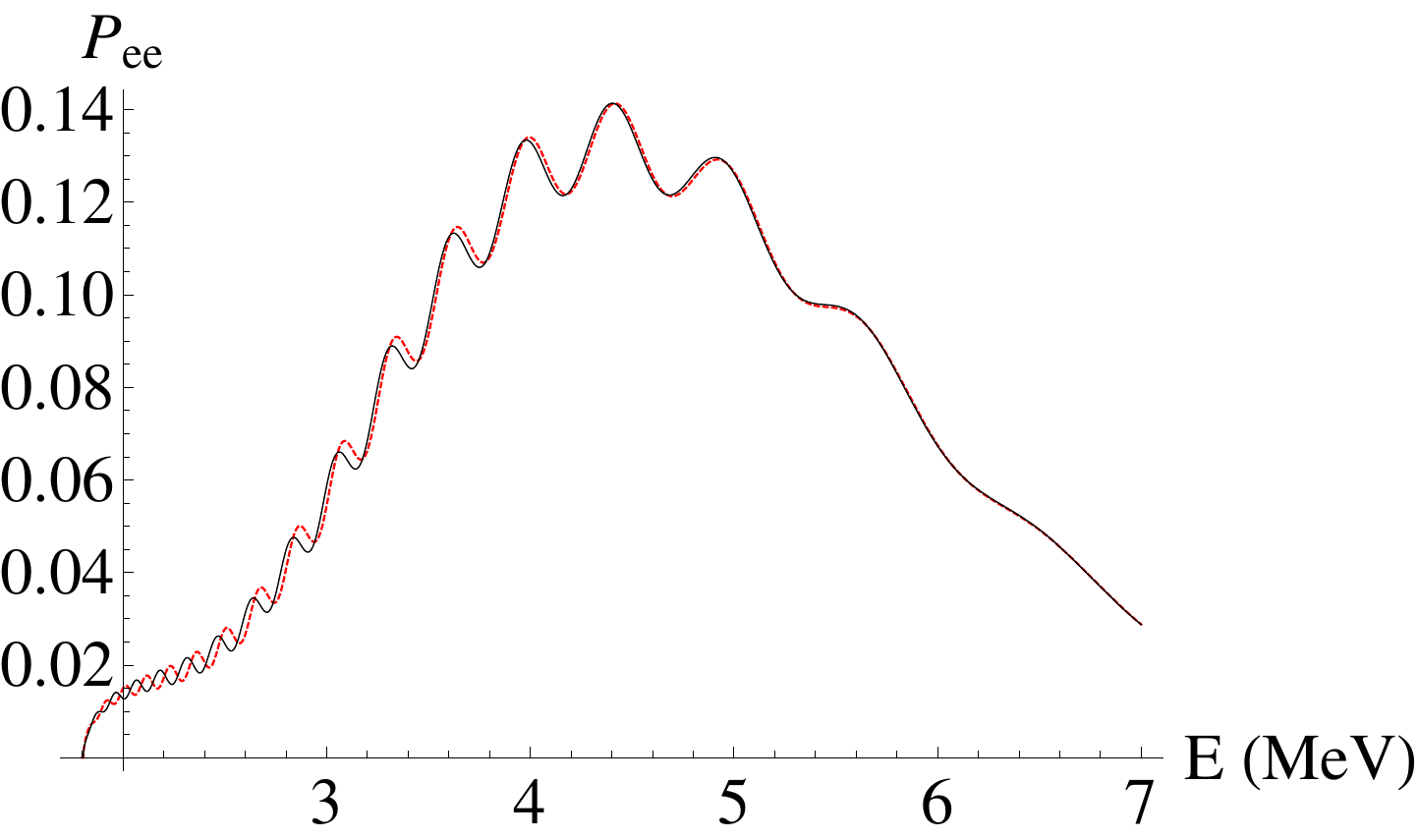}
\includegraphics[width=3.2in,height=2in]{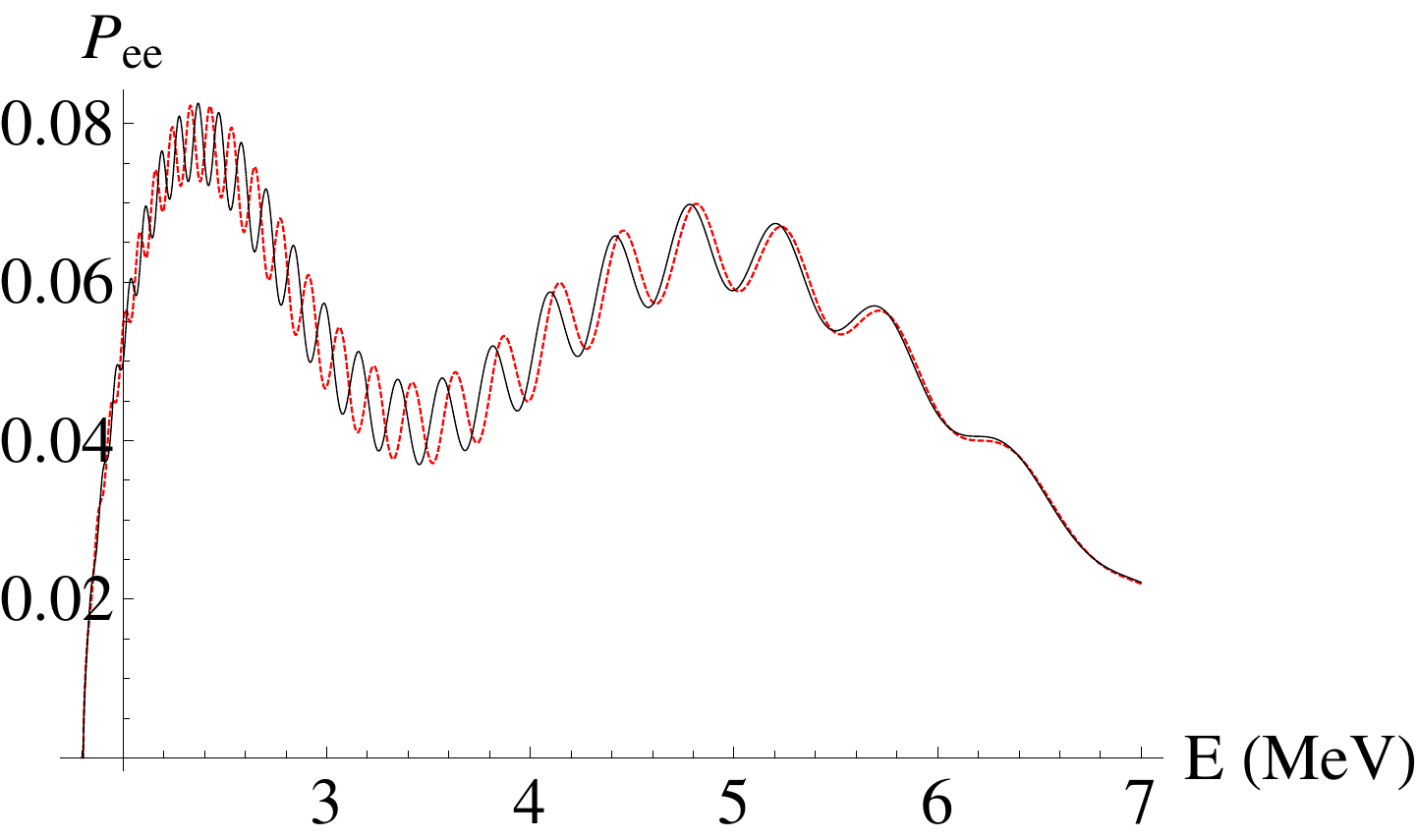}
\caption{Theoretical neutrino fluxes, including 3 flavor oscillation, at 40 and 58 km for both hierarchies with the same value of $\Delta M^2_{eff}=2.46\times 10^{-3}\ \mathrm{eV}^2$.  In the left panel, at 40 km, the hierarchies are difficult to distinguish because $\an$ is nearly linear at the visible peaks.  This degeneracy is broken by the higher $n$ peaks visible at 58 km, seen in the right panel.}
\label{degenfig}
\end{center}
\end{figure}

Thus if the baseline is short enough so that only these peaks may be reliably measured, then there will be a value of $\Delta M^2_{eff}$ that reproduces the peaks for both hierarchies, as seen at 40 km in the first panel of Fig. \ref{degenfig}.   Here the peak $n=6$ can barely be seen at 6.6\ MeV, whereas $n=7$ at 5.7 MeV is clearly discernible.  The largest peaks are $n=8,\ 9,\ 10,$ and $11$.  However the energies of these peaks are independent of the hierarchy at constant $\Delta M^2_{eff}$.  The hierarchy-dependence becomes somewhat larger at the 12th peak, which is located at about 3.32 MeV in the case of the normal hierarchy and 3.30 MeV in the case of the inverted hierarchy.  Thus if $\Delta M^2_{eff}$  peaks can be determined at better than 1\% from the low $n$ peaks and then the location of the 12th peak can be determined with a precision of better than 1\%, a determination of the hierarchy would be barely possible.  The lower energy peaks are smaller, but more hierarchy dependent.  For example, the 16th peak would be at 2.50 MeV with the normal hierarchy, but 2.46 MeV with the inverted hierarchy.  This 2\% difference is just within the resolution of the proposed detectors of Refs. \cite{caojun,caojunseminario,yifangseminario}, and so when combined with an accurate measurement of $\Delta M^2_{eff}$ one could potentially determine the hierarchy at the 1-2$\sigma$ level.



In the second panel one can see the electron neutrino survival probability at 58 km with both hierarchies and the same value of $\Delta M^2_{eff}$.  At this long baseline one can see maxima up to $n=20$, where the nonlinearity in $\an$ is appreciable and so, as one can see, the low and mid energy peaks are hierarchy-dependent at the 1-2\% level.  

\subsection{The 1-2 oscillation maximum}
It is clear from Fig. \ref{degenfig} that if $\Delta M^2_{eff}$ is determined from the low $n$ peaks then the locations of the peaks at the 1-2 oscillation maximum
\beq
n=\frac{1}{2\epsilon}\sim 16\hsp
E=\frac{L}{18\mathrm{\ km}}\textrm{MeV}
\eeq
are hierarchy-dependent, and so one may hope to use their locations to determine the hierarchy.  This can be done if the peak locations allow for a combination of the mass differences which is distinct from $\Delta M^2_{eff}$.  In fact, two such combinations can be determined, one from the location of the peaks and one from the distance between them.

To derive these two combinations, we will need to find $\an$  near the 1-2 oscillation maximum.  This is easily obtained by expanding (\ref{alfeq}) about $n=1/2\epsilon$
\beq
0\sim \sin(2\pi\an)-\tp212\sin\left(2\pi\left[\an-\epsilon \left(n-\frac{1}{2\epsilon}\right)\right]\right).
\eeq
Linearly expanding the sine function yields
\beq
0\sim (1-\tp212)\an+\epsilon \left(n-\frac{1}{2\epsilon}\right)
\eeq
and so for $n\sim 1/2\epsilon$
\beq
\an\sim-\epsilon \frac{n-\frac{1}{2\epsilon}}{ 1-\tp212}.  \label{anminimo}
\eeq

At $n=1/2\epsilon$, $\an=0$.  Of course $n$ must be an integer, so it can never be precisely equal to $1/2\epsilon$.  Nonetheless, $\an$ will be of order $0.01$ when $n$ is at the closest integral value, leading to a less than 0.1\% contribution to the energy.  When $\an=0$, the energy of the peak is 
\beq
E=\frac{4\pi n L}{\mn31}=\frac{2\pi L}{\m21}
\eeq
allowing for a precise determination of $\m21$.  Of course, to know that one is at the 1-2 maximum by counting peaks, one must know $\epsilon$ and so this is related to a measurement of $\mn31$.  Alternately, one can find the 1-2 maximum by looking at the large oscillations in the flux due to 1-2 mixing and then count peaks to determine $n$ at the minimum, which allows for a determination of $\epsilon$ and so $\mn31$ from $\m21$.

The distance between the peaks near the 1-2 maximum allows for a determination of the mass differences which is independent of precise knowledge of the 1-2 oscillation parameters.  Inserting the linear expansion of $\an$\ (\ref{anminimo}) into the definition of $\an$\ (\ref{pichi}) one finds that the distance between two peaks is
\beq
\Delta\left(\frac{L}{E}\right)=\frac{4\pi}{\mn31}\left(1\mp\frac{\epsilon}{1-\tp212}\right)
\eeq
and so it determines the effective mass
\beq
\Delta M^2_{min}=\left(1\pm\frac{\epsilon}{1-\tp212}\right)\mn31=\frac{2-\tp212}{1-\tp212}\mn31-\frac{1}{1-\tp212}\mn32 .
\eeq 

This effective mass is quite different from that defined in Eq. (\ref{meff}).  Approximating $\tp212=1/2$ one finds that the spacing between the first 10 peaks yields an effective mass
\beq
\Delta M^2_{eff}=\frac{2}{3}\mn31+\frac{1}{3}\mn32 \label{meffdue}
\eeq
while the peak spacing between $n=14$ and $n=18$ yields
\beq
\Delta M^2_{min}=3\mn31-2\mn32 . \label{mmindue}
\eeq
A detector at a baseline of less than 50 km can accurately determine the combination (\ref{meffdue}) while one with a baseline between 45 and 70 km, as it sees higher $n$ peaks, may more easily measure the combination (\ref{mmindue}).   The normal mass hierarchy is equivalent to the second mass being larger than the first, and so by comparing these masses one can determine the hierarchy.  

The difference between these masses is quite large, about 7\%, however a 7\% measurement of $\Delta M^2_{min}$ requires a 7\% precision in the measurement of the difference between the the peaks in the linear regime near to the 1-2 oscillation maximum $n=1/2\epsilon$.   As can be seen in Fig \ref{anfig}, this linear regime includes about 8 peaks, corresponding to a 40\% variation in energy.  This means that a 7\% precision in the distance between the peaks requires a 4\% precision in the energy of the peaks, much less than was required at shorter baselines.  Therefore a comparison of the low $n$ peak positions and the 1-2 maximum peak separations is a promising test of the hierarchy, so long as the statistics are sufficient for the peaks to be observed.

\section{Resolution and flux constraints} \label{vincsez}

\subsection{Energy resolution} \label{ressez}

\begin{figure} 
\begin{center}
\includegraphics[width=3.2in,height=2in]{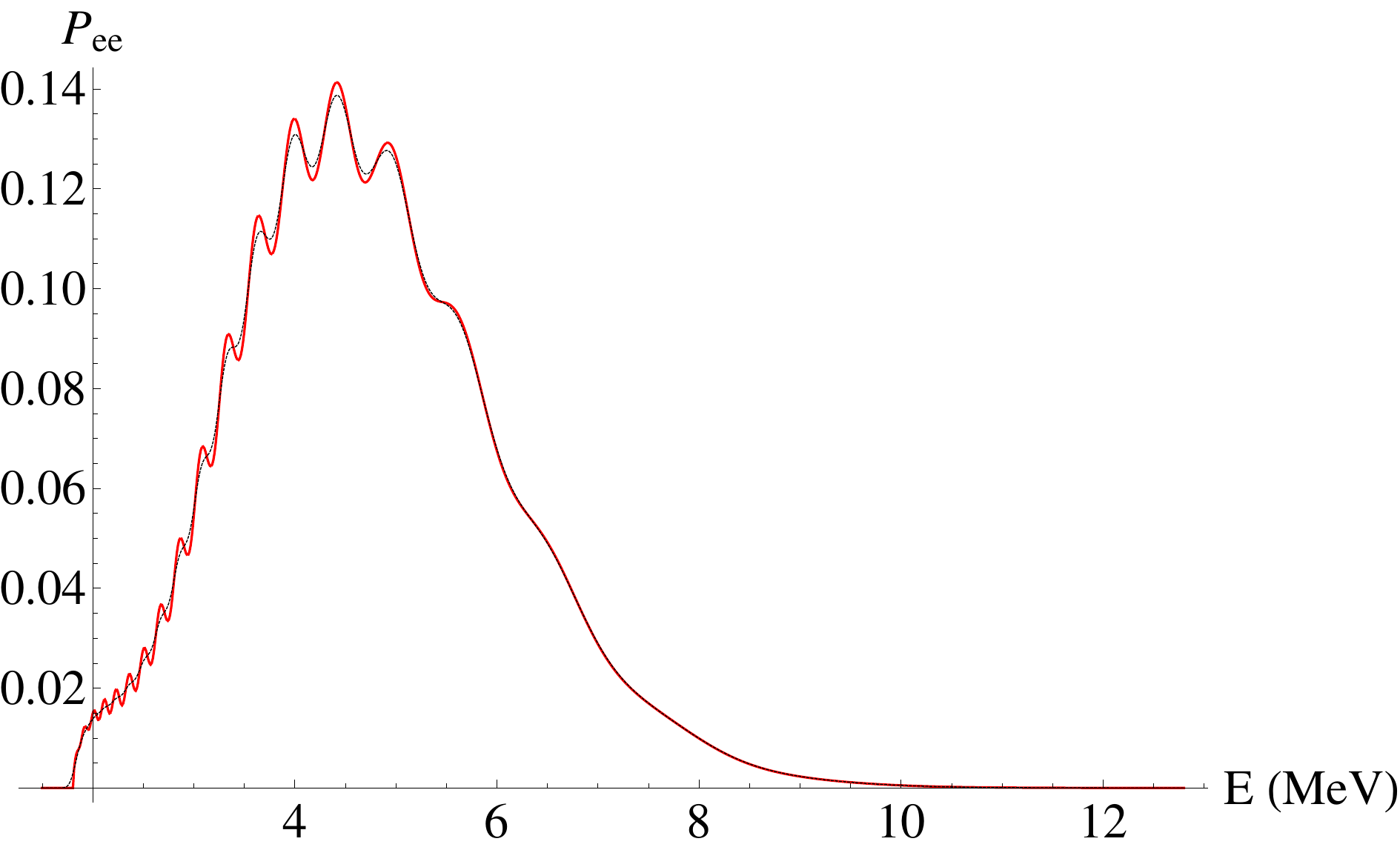}
\includegraphics[width=3.2in,height=2in]{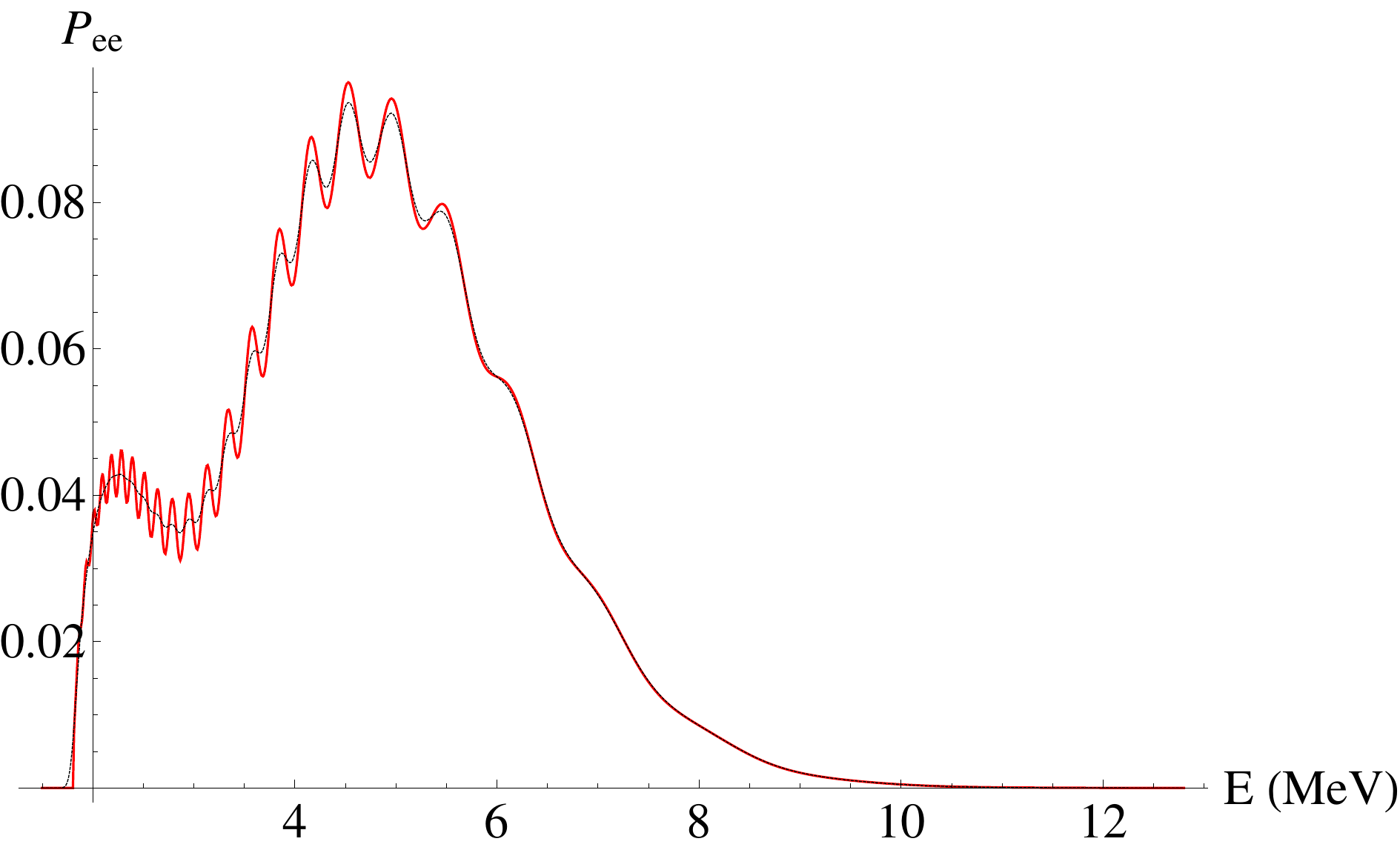}
\includegraphics[width=3.2in,height=2in]{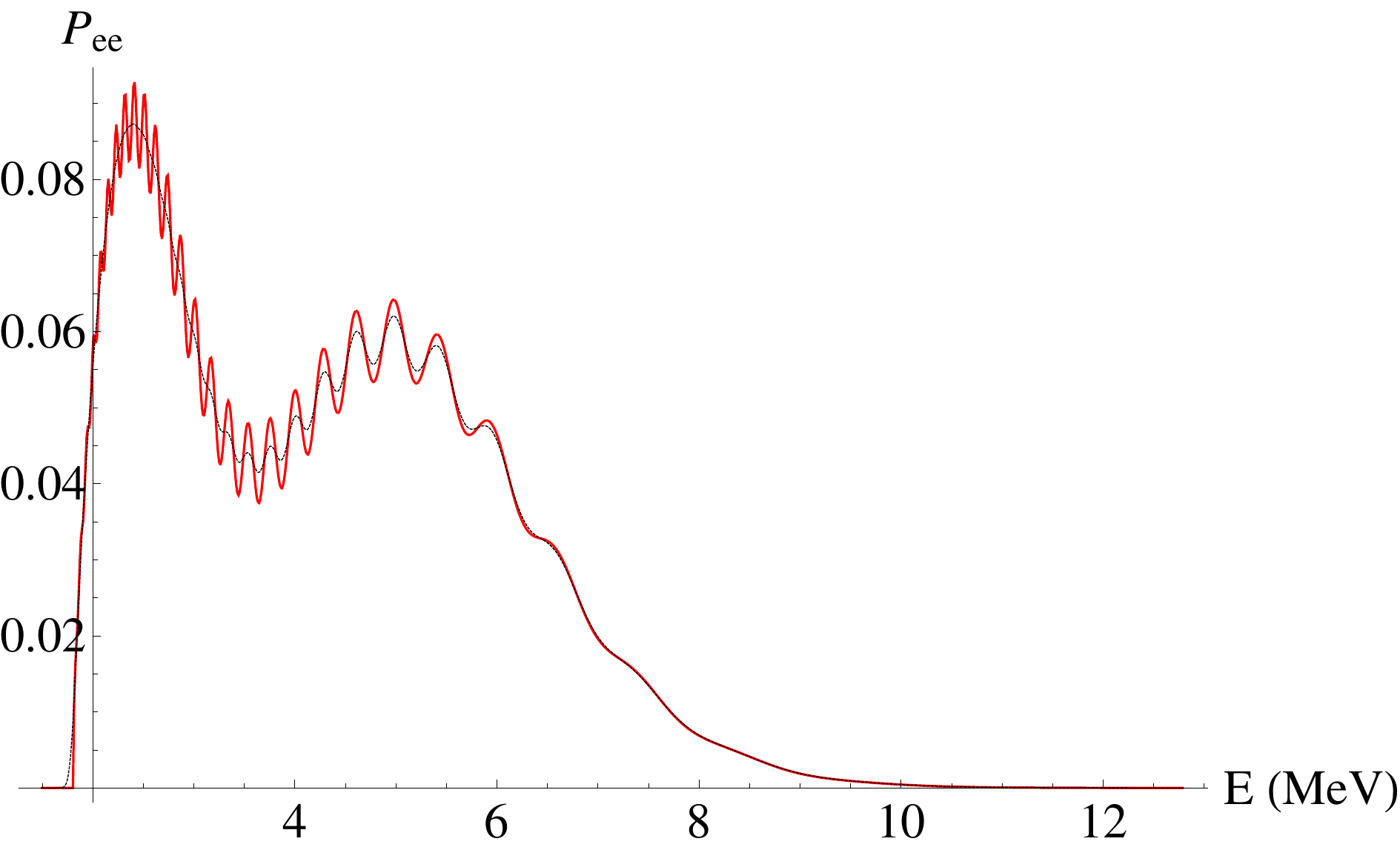}
\includegraphics[width=3.2in,height=2in]{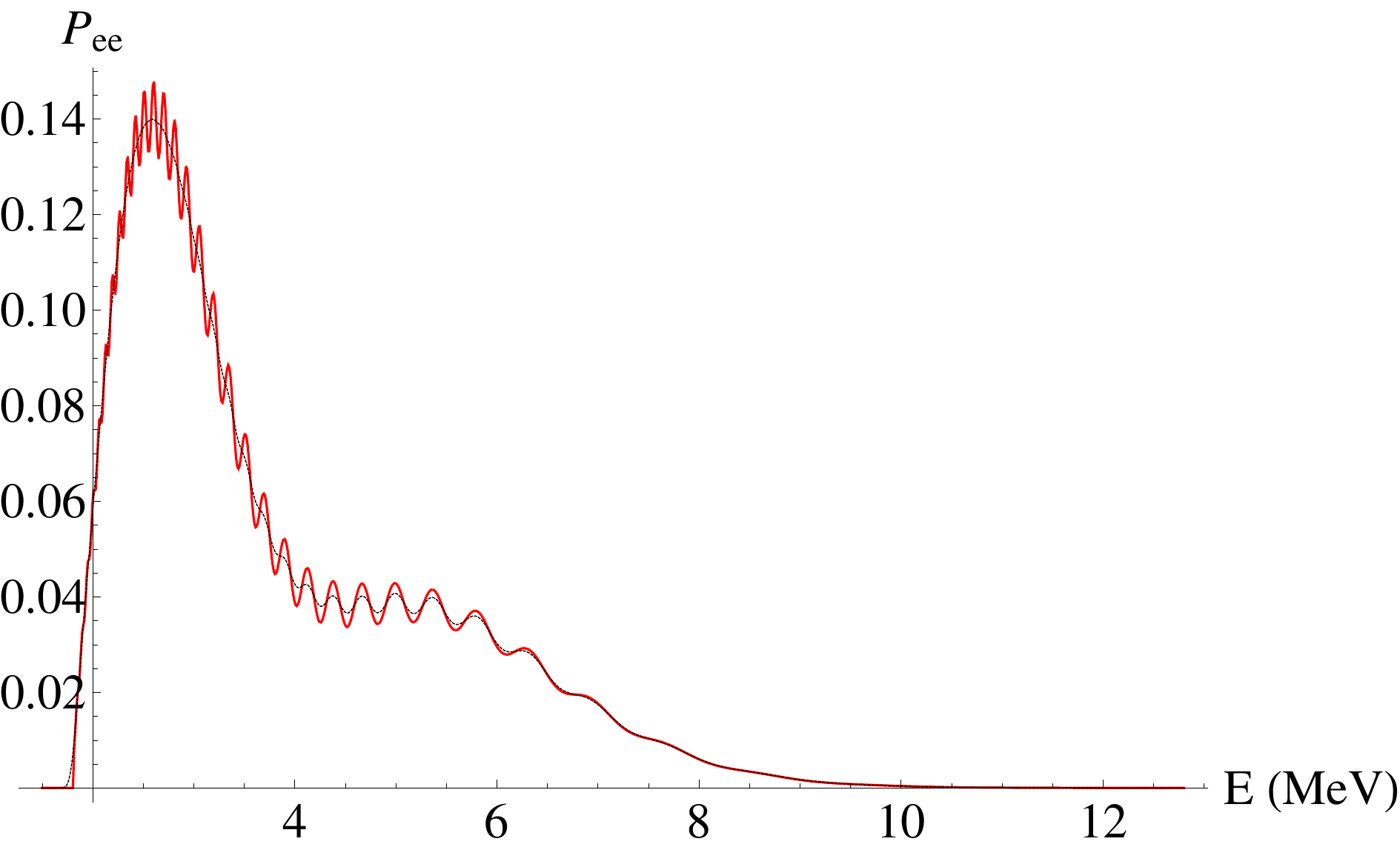}
\caption{The true neutrino spectrum $P$ and the observed spectrum $P_{\rm{obs}}$ with a resolution of $3\%/\sqrt{E}$ using the normal hierarchy at baselines of 40, 50, 60 and 70 km.  Notice that the lowest energy oscillations are smeared away and the energy threshold for this smearing increases with the baseline, such that at higher baselines the maximum $n$ observable increases slowly.}
\label{smearfig}
\end{center}
\end{figure}

The neutrino energy $E$ is determined from the positron energy $E_e$ by adding 780 keV (\ref{posenergia}).   The positron energy is determined by counting photoelectrons in a scintillator.  The number of photoelectrons is proportional to the positron energy, and so the energy resolution $\sigma_E$ is proportional to the square root of the positron energy.  For example, at Daya Bay II it has been suggested in Ref. \cite{caojun} that a resolution of
\beq
\sigma_E=0.03\sqrt{(E_e){\mathrm{MeV}}}
\eeq
will be achieved, corresponding to a fractional resolution of $3\%/\sqrt{E/\mathrm{MeV}}$.  We will assume this resolution in what follows.  The observed positron  energy spectrum is then the convolution of the true spectrum with a Gaussian of width $\sigma$
\beq
P_{\mathrm{obs}}(E_e)=\int dE^{\rm{true}}_e P(E_e^{\rm{true}}) e^{-\frac{(E_e-E_e^{\textrm{true}})^2}{2\sigma_E^2}}.
\eeq
Both spectra are displayed in Fig. \ref{smearfig}.

The observed spectrum $P_{\mathrm{obs}}(E_e)$ is the best that can be hoped for, once the response of the detector is understood and with an infinite number of neutrinos.  Finite flux effects will be briefly discussed in Subsec. \ref{flussosez} and then discussed in detail in the companion paper on our simulation results.  A poorly understood nonlinear detector response is fatal to the Fourier analysis that will be discussed in Sec. \ref{fouriersez}, but individual peaks may be analyzed where the response is understood.

Even in this idealized setting, it is clear from Fig. \ref{smearfig} that the low energy (high $n$) peaks cannot be resolved.  How high is $n$ for the lowest energy peak that can be resolved?  This depends on the baseline and the neutrino flux.  However a rough answer is obtained by asserting that the distance between the $n$th maximum and the adjacent minimum (\ref{corto}) 
\beq
\Delta E=0.97\frac{L/\mathrm{km}}{n}\mathrm{MeV}-0.97\frac{L/\mathrm{km}}{n+1/2}\mathrm{MeV}= 0.49 \frac{L/\mathrm{km}}{n^2}\mathrm{MeV}
\eeq
be greater than
\beq
2\sigma =0.06\sqrt{(E_e){\mathrm{MeV}}}=0.06\mathrm{\ MeV}\sqrt{0.97\frac{L/\mathrm{km}}{n}-0.78}.
\eeq
These two quantities are equal when 
\beq
L/\mathrm{km}=7.4\times 10^{-3}n^3\left(1\pm\sqrt{1-\frac{223}{n^2}}\right). \label{lnecc}
\eeq
The equality with the plus sign yields the upper bound on observable $n$ for a given baseline $L$, whereas the expression with the minus sign typically yields a value of $n$ at energies where no reactor neutrinos are observed.  

Alternately (\ref{lnecc}) provides the baseline $L$ necessary to observe the $n$th peak.  When $n\leq 15$ it provides no bound at all, so long as there is enough flux and the detector response is well understood, the energy resolution is not an obstruction to observing these peaks.  Of course at low baselines they may be difficult to observe because there simply are not many or any neutrinos observed at the corresponding energy.  At $n=16$, $17$ and $18$\ one finds minimum baselines of 45 km, 57 km and 70 km respectively.  In practice the true minimum depends on the neutrino flux.  But this rough estimate shows an essential point, that with a $3\%/\sqrt{E}$ fractional resolution any medium baseline, between about 40 and 70 km, is sufficient to observe the 1-2 oscillation maximum if there is enough neutrino flux.  Longer baselines only marginally extend the reach to higher peaks, although each of these peaks is in the 1-2 maximum region and so even just 1 or 2 more peaks can greatly enhance the possibility of determining the mass hierarchy.   

While the maximum observable $n$ is reasonably independent of the baseline, this derivation shows that it is strongly dependent upon $\sigma_E$.  A resolution comparable to that of Daya Bay or RENO would lead to a maximum $n$ which is still in the linear region of $\an$, and so the hierarchy would be unobservable.

\subsection{How much neutrino flux is required?} \label{flussosez}

The neutrino flux that can be observed by a large, distant detector via inverse $\beta$ decay is still quite unknown, even the efficiency of the detector is difficult to predict.  A rough estimate can be made using the flux observed at Daya Bay, using the flux normalization-independent determination of $\theta_{13}$ to eliminate the loss due to 1-3 operation.  At Daya Bay 17.4 GW of thermal power yields 80 neutrinos/day at each 20 ton detector at a weighted baseline of 1600 meters.  Therefore the total measured flux/year from a $P$ GW reactor complex measured at a detector of mass $M$ at a baseline $L$ is roughly
\beq
\Phi=365\times 80\times \frac{P}{17.4\ \mathrm{GW}}\frac{M}{0.02\ \mathrm{ktons}}\frac{2.6\ \mathrm{km}^2}{L^2}=2.2\times 10^5\frac{(P/\mathrm{GW})(M/\mathrm{ktons})}{(L/\mathrm{km})^2}. \label{totflusso}
\eeq
For example, at a distance $L$ from the Daya Bay and Ling Ao reactors a 20 kton detector may observe $7.6\times 10^7/L^2$ neutrinos/year, where the length $L$ is measured in kilometers.

Let the energy width of a peak be $\Delta E$, the integrated flux within that energy range be $\phi$ and the peak be a fraction $A$ higher than the flux in that energy range from the reactor in question considering 1-2 oscillations and also the flux from distant reactors.   In the absence of other reactors the fraction $A$ varies from $\spp2213$ at low $n$ to $1.7\spp2213$ at the 1-2 maximum.  The fractional error in the flux in that range will be $1/\sqrt{\phi}$.  Therefore the peak may be observed if $\sqrt{\phi} A>1$.  

Simply observing the peaks is useful for two reasons.  First of all, the distance between peaks is roughly $L\mn31/\pi$ and so by identifying peaks, one can check the consistency of the location of other peaks.  Second, recall that it is easier to determine the $n$ value of the well-separated, high energy, low $n$ peaks.  If one can observe the peaks from low to high $n$ then it is possible to count them and so determine the $n$ values of the low energy peaks.  While the hierarchy can be determined from the distance between the high $n$ peaks, as described in the previous subsection, without knowing the precise value of $n$, nonetheless if one knows the $n$ value of a high $n$ peak it can be used to determine a combination of the mass differences via (\ref{pichi}).  This determination has a precision of $1/n$, and so at high $n$ it leads to a precise determination.

However, to determine the hierarchy it is not enough to observe the peaks, one must determine their energies as precisely as possible.  How precisely may they be determined?  They may be determined within an energy $\delta E$ if the neutrino surplus can be seen in width $\delta E$ bands within the peak.  This requires
\beq
1<\sqrt{\phi}\frac{A\delta E}{2\Delta E}=\sqrt{\phi}Afn
\eeq
where $f$ is the fractional energy precision desired.  This implies that the maximal fractional precision with which the energy of a peak can be measured by a detector with perfect resolution is
\beq
f>\frac{1}{An\sqrt{\phi}}.
\eeq
As mentioned above, with no unwanted backgrounds from other reactors, $A$ varies between $0.09$ at the 1-2 maxima to $0.16$ at the minima.  

Consider for example the 11th peak at a 40 km baseline, which lies at 3.6 MeV.  At this point $A\sim 0.12$.  The flux within the peak is about 4\% of the total flux (\ref{totflusso}), which each year at a 20 kton detector at 40 km from Daya Bay may be
\beq
\phi=0.04\times 7.6\times 10^7/(40)^2=1.9\times 10^3.
\eeq
Therefore, after $m$ years, the best resolution of an ideal detector would be
\beq
f_{min}=\frac{1}{0.12\times 10\sqrt{1.9m\times 10^3}}=\frac{1}{57\sqrt{m}}.
\eeq
Thus an ideal detector can find the 11th peak energy to within $1/\sqrt{m}$ percent after $2.5m$ years.  As was described in Subsec. \ref{ressez}, since the peak width is greater than the high energy resolution proposed at a new medium baseline detector experiment in Ref. \cite{caojun}, resolution effects will not significantly alter the determination of this peak.  Therefore one may expect $\Delta M^2_{eff}$ to be determined to a precision  better than 1 percent at a 40 kilometer baseline experiment with no backgrounds from other reactors.

However, to determine the hierarchy, one also needs to determine another combination of the neutrino mass differences.  This requires the measurement of a higher $n$ peak.  Consider for example the $n=16$ peak at 2.5 MeV.  While it is in general a poor approximation to ignore the backgrounds provided by distant reactors, if one does ignore them then, since the peak is near the 1-2 maximum, the relative peak height is $A=0.16$.   The total flux in the peak is only about 0.8\% of the total, and so 
\beq
f_{min}=\frac{1}{0.16\times 16\sqrt{3.8m\times 10^2}}=\frac{1}{53\sqrt{m}}.
\eeq
As can be seen in Fig \ref{degenfig}, a 3\% precision is sufficient to determine the hierarchy and so the energy of this peak would yield a $2\sigma$ determination after 2 to 3 years.  As the width of the peak is about 3\%, a $3\%/\sqrt{E}$ resolution may reduce the amplitude by a factor of 2, still allowing for a determination of the hierarchy.

\section{The Fourier transform of the survival probability} \label{fouriersez}

While each peak provides some information regarding a combination of neutrino mass differences and therefore the hierarchy, it may well be that the fluxes are too weak or the backgrounds too large for the peaks to be reasonably well identified.  Complimentary information can be obtained by combining the peaks.  As the electron survival probability $P_{ee}$ is a sum of periodic cosine functions, they can be combined by a Fourier transform.  Even when individual peaks are hard to identify, the combination probed by the Fourier transform may well be visible and so may provide the best chance for determining the hierarchy \cite{hawaii}.

\subsection{The complex Fourier transform}

For simplicity we will approximate the observed electron antineutrino spectrum by a Gaussian distribution in $L/E$ space
\beq
\Phi\left(\frac{L}{E}\right)=e^{-\left(\frac{L}{E}-L\langle\frac{1}{E}\rangle\right)^2/\sigma^2} \label{spec}
\eeq
where $\langle 1/E \rangle$ is the average $1/E$ of a detected neutrino.  While with only slightly more complicated equations the following could be avoided, we will make the crude approximation that (\ref{spec}) is the neutrino spectrum after 1-2 neutrino oscillations, and that the expectation value of the inverse energy is therefore taken with respect to the 1-2 oscillated spectrum, which depends upon $L$. 

1-3 oscillations affect this spectrum by introducing a modulation equal to $\Phi(L/E) P_{13}(L/E)$.  The Fourier transform of this modulation is
\bea
F_{13}(k)&=&\int d\left(\frac{L}{E}\right) \Phi(L/E)  P_{13}(L/E) e^{i\frac{kL}{E}}\\&=&\frac{\sigma\sqrt{\pi}\cp212}{4}\left(e^{-\frac{\sigma^2}{4}\left(k+\frac{\mn31}{2}\right)^2}e^{i\left(k+\frac{\mn31}{2}\right)L\langle\frac{1}{E}\rangle}+(e^{-\frac{\sigma^2}{4}\left(k-\frac{\mn31}{2}\right)^2}e^{i\left(k-\frac{\mn31}{2}\right)L\langle\frac{1}{E}\rangle}\right)\nonumber
\eea
where we have factored the $\spp2213$ out of the definition of $F_{13}$. This quantity has two peaks, one at $k=\mn31/2$ and one at $k=-\mn31/2$.  At each peak, one of the two terms in the parenthesis dominates, and the other is suppressed by a factor of order $e^{n^2/4}$, which is large enough that we will neglect the subdominant term.  Therefore, near the first peak
\beq
F_{13}(k)=\frac{\sigma\sqrt{\pi}\cp212}{4}e^{-\frac{\sigma^2}{4}\left(k-\frac{\mn31}{2}\right)^2}e^{i\left(k-\frac{\mn31}{2}\right)L\langle\frac{1}{E}\rangle}.
\eeq
The norm of the complex function $F_{13}$ has a maximum at $k=\mn31/2$, where $F_{13}$ is real.  The real part of $F_{13}$, corresponding to a cosine transform, is, within the validity of the approximations described above, symmetric about this maximum.  The imaginary part, corresponding to a sine transform, vanishes at this maximum and is antisymmetric about it.

The transform $F_{23}(k)$ can be calculated identically.  Near the positive $k$ maximum the sum of the two transforms is just
\bea
F(k)&=&\frac{\sigma\sqrt{\pi}}{4}\left[\cp212 e^{-\frac{\sigma^2}{4}\left(k-\frac{\mn31}{2}\right)^2}e^{i\left(k-\frac{\mn31}{2}\right)L\langle\frac{1}{E}\rangle}+\sp212 e^{-\frac{\sigma^2}{4}\left(k-\frac{\mn32}{2}\right)^2}e^{i\left(k-\frac{\mn32}{2}\right)L\langle\frac{1}{E}\rangle}\right]\nonumber\\
&=&\frac{\sigma\sqrt{\pi}}{4}\left[\cp212 e^{-\frac{\sigma^2}{4}\left(k-\frac{\mn31}{2}\right)^2}+\sp212 e^{-\frac{\sigma^2}{4}\left(k-\frac{\mn32}{2}\right)^2}e^{\pm i\frac{\m21 L}{2}\langle\frac{1}{E}\rangle}\right]e^{i\left(k-\frac{\mn31}{2}\right)L\langle\frac{1}{E}\rangle} \label{xform}
\eea
where the $+$ sign applies to the normal hierarchy.

The first term corresponds to the old peak, at $k=\mn31/2$ and the second to a new peak at $k=\mn32/2$, which on its own would have a height equal to $\tp212\sim 1/2$ times that of the first.  Of course, due to the phase difference between the two terms, for some choice of parameters the interference between the two terms implies that the second is not a local maximum of the norm of the Fourier transform.  The $k$ value of the peak then gives the corresponding mass difference.  Thus if the smaller peak is to the left, corresponding to smaller $k$, of the larger peak then $\mn32<\mn31$ and so one can conclude that there is a normal neutrino mass hierarchy \cite{hawaii}.

\subsection{Determining $\mn31$ at the 1-2 oscillation maximum}

We will refer to the baseline
\beq
L=\frac{2\pi}{\m21\langle\frac{1}{E}\rangle}
\eeq
as the 1-2 oscillation maximum, as it is roughly the baseline at which the largest fraction of neutrinos disappears as a result of $P_{12}$.   It is approximately 58 km.  At this distance the Fourier transform (\ref{xform}) simplifies as the two terms are precisely out of phase
\beq
F(k)=\frac{\sigma\sqrt{\pi}}{4}\left[\cp212 e^{-\frac{\sigma^2}{4}\left(k-\frac{\mn31}{2}\right)^2}-\sp212 e^{-\frac{\sigma^2}{4}\left(k-\frac{\mn32}{2}\right)^2}\right]e^{i\left(k-\frac{\mn31}{2}\right)\frac{2\pi}{\m21}} . \label{segno}
\eeq

The cosine transform is just the real part of $F$.  Up to $k$-independent factors, it is proportional to 
\beq
F_{\cos}(\tilde{k})=\left( e^{-\frac{\sigma^2}{4}\tilde{k}^2}-\tp212  e^{-\frac{\sigma^2}{4}\left(\tilde{k}\pm\frac{\m21}{2}\right)^2}\right)\cos\left(\frac{2\pi\tilde{k}}{\m21}\right) \label{coseq}
\eeq
where we have defined the distance from the $1-3$ peak to be
\beq
\tilde{k}=k-\frac{\mn31}{2}.
\eeq
The maxima of $F_{\cos}$ are found by setting to zero its derivative with respect to $\tilde{k}$, yielding the condition
\bea
&&\left(\frac{\sigma^2}{2}\tilde{k}+\frac{2\pi}{\m21}\tan\left(\frac{2\pi\tilde{k}}{\m21}\right)\right)e^{-\frac{\sigma^2}{4}\tilde{k}^2}\\&&\ \ \ \ \ \ \ \ \ \ \ \ \ =
\tp212\left(\frac{\sigma^2}{2}\left(\tilde{k}\pm\frac{\m21}{2}\right)+\frac{2\pi}{\m21}\tan\left(\frac{2\pi\tilde{k}}{\m21}\right)\right)e^{-\frac{\sigma^2}{4}\left(\tilde{k}\pm\frac{\m21}{2}\right)^2}\nonumber
\eea
where again the $+$ sign corresponds to the normal hierarchy and the $-$ sign to the inverted hierarchy.  The largest peak is the closest to $\tilde{k}=0$, the absolute maximum of the Fourier transform of $P_{13}$.  To find this peak we may expand $\tilde{k}$ about 0, at linear order we find
\beq
\tilde{k}=\pm\frac{\m21/2}{1+\frac{\pi^2}{2\beta}\left(e^\beta\ctp212-1\right)+2\beta} \label{ktilde}
\eeq
where we have defined the constant
\beq
\beta=\frac{\sigma^2\left(\m21\right)^2}{16}.
\eeq

As the denominator of (\ref{ktilde}) is much greater than 1, we learn that
\beq
|\tilde{k}|<<\frac{\m21}{2}
\eeq
and so the maximum of $F_{\cos}$ lies at approximately
\beq
k_{max}=\frac{\mn31}{2}.
\eeq
This means that if the detector is placed at the 1-2 oscillation maximum baseline then the peak of the cosine transform of the full neutrino spectrum lies essentially at the peak of $P_{13}$ alone.  This is not because because the frequencies of the $P_{13}$ and $P_{23}$ terms are similar, but because as seen in Fig. \ref{cosfig}, at this baseline the absolute maximum of the Fourier transform of $P_{13}$, which corresponds to its frequency, happens to coincide with one of the minima of the cosine transform of $P_{23}$, which is not its frequency.  {\textbf{Thus, at the 1-2 oscillation maximum, the absolute maximum of the sum of two cosine transforms is coincident with that of $P_{13}$ alone, allowing for a direct and precise determination of $\mn31$}}.  This may be useful on its own even if the hierarchy has already been determined by NOvA or T2K.

\subsection{Determining the hierarchy at the 1-2 oscillation maximum}

Can this simplification also yield information about the hierarchy?  A precise determination of $\mn31$ combined with MINOS' best fit for $\mn32$ could give a 1$\sigma$ answer to this question, within 5 years MINOS+ may improve this to 2$\sigma$.  But with enough flux the peak structure of the Fourier transform alone yields some information about the hierarchy.  

\begin{figure} 
\begin{center}
\includegraphics[width=5.2in,height=2.5in]{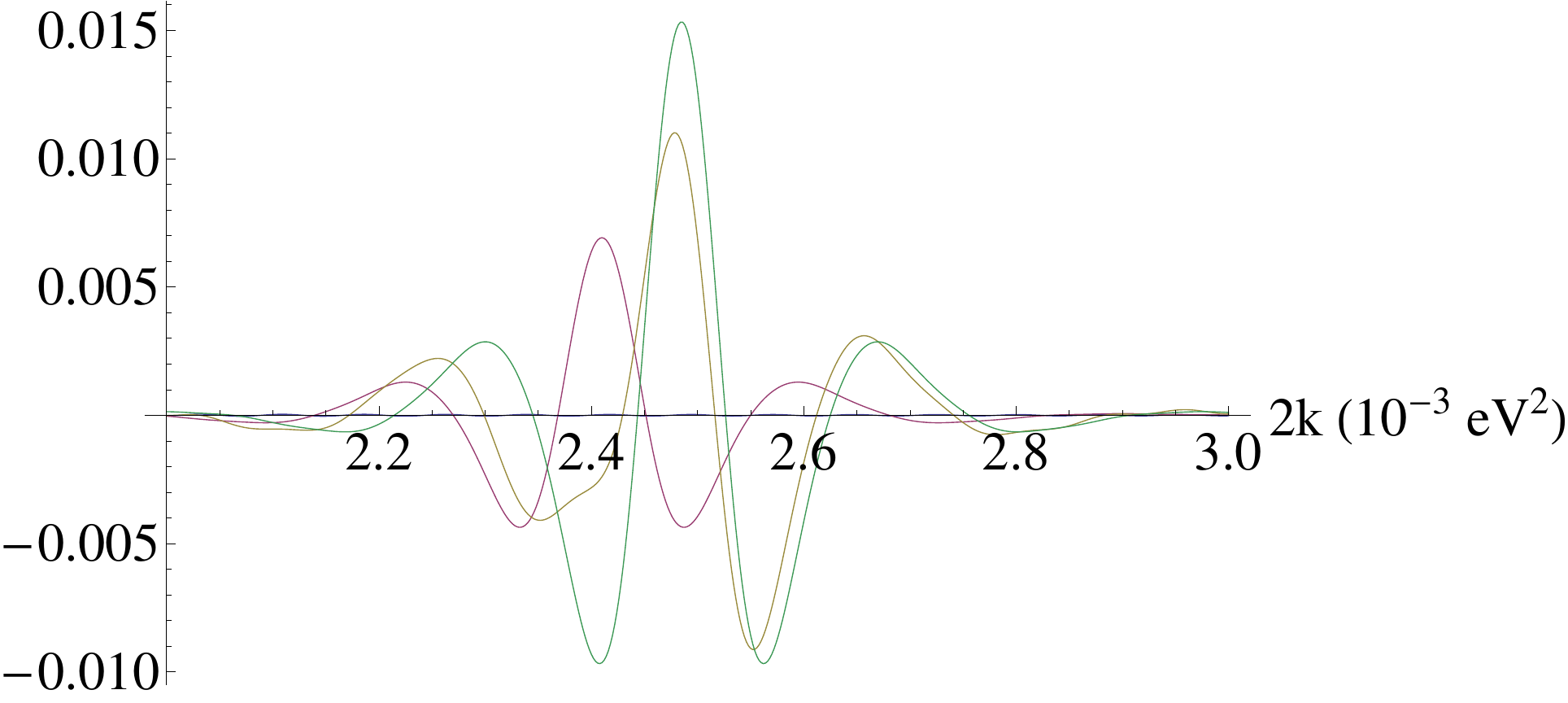}
\caption{The cosine Fourier transform of $P_{12}$ (blue), $P_{23}$ (purple), $P_{13}$ (green) and $P_{ee}$ (yellow) at 58 km in the case of the normal hierarchy.  Note that the $P_{13}$ and $P_{23}$ curves are one half of a wavelength out of phase, so that the total extrema coincide with those of $P_{13}$.}
\label{cosfig}
\end{center}
\end{figure}

As can be seen in Fig. \ref{cosfig}, just as the relative sign in (\ref{segno}) implied that the global maximum of the cosine transform of the total spectrum is essentially coincident with that of $P_{13}$, it also implies that the minima just to its left and right are roughly coincident with those of $P_{13}$
\beq
k_{\min}^L=\frac{\mn31-\m21}{2}\hsp
k_{\min}^R=\frac{\mn31+\m21}{2}.
\eeq
These are minima for the simple reason that the cosine on the right of (\ref{coseq}) is equal to $-1$.  Substituting these values of $k$ into (\ref{coseq}) we find that the values of the cosine transforms at the two minima are
\beq
F_{\cos}(k_{min}^L)=-\left(e^{-\beta}-\tp212 e^{-\beta(-1\pm 1)^2}\right)\hsp
F_{\cos}(k_{min}^R)=-\left(e^{-\beta}-\tp212 e^{-\beta(1\pm 1)^2}\right)
\eeq
where as usual the upper sign corresponds to the normal hierarchy.  In the case of the normal hierarchy, the positive sign means that the second term is larger on the left, which implies that the minimum on the right is deeper.  In the case of the inverted hierarchy these two depths are interchanged, and so the peak on the left is deeper.  Thus we have recovered the criterion for determining the hierarchy from the cosine transform which was proposed in Ref. \cite{caojun}.

\begin{figure} 
\begin{center}
\includegraphics[width=5.2in,height=2.5in]{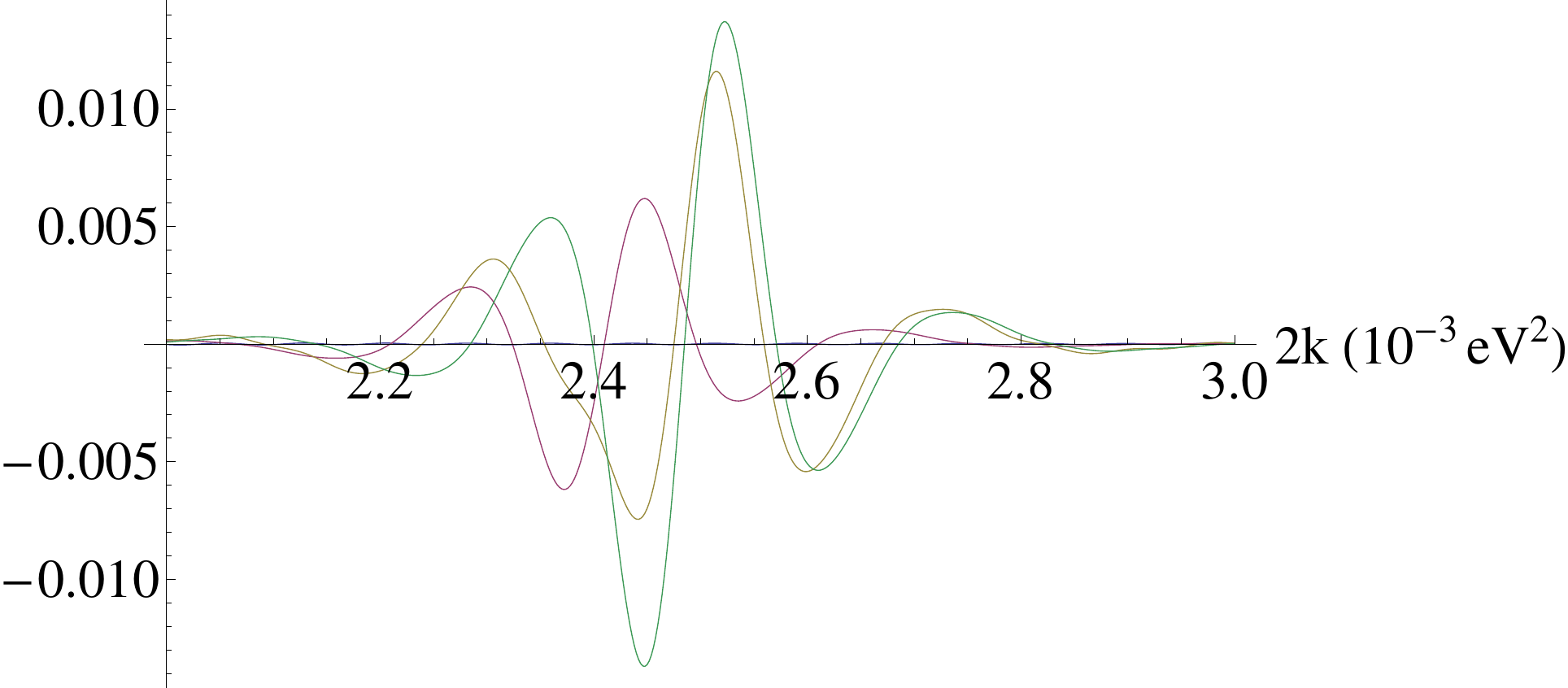}
\caption{The sine Fourier transform of $P_{12}$ (blue), $P_{23}$ (purple), $P_{13}$ (green) and $P_{ee}$ (yellow) at 58 km in the case of the normal hierarchy. Note that the $P_{13}$ and $P_{23}$ curves are one half of a wavelength out of phase, so that the total extrema coincide with those of $P_{13}$.}
\label{sinfig}
\end{center}
\end{figure}

A similar analysis can be applied to the imaginary part of $F(k)$, which is obtained via a sine transform.  As $F(\mn31/2)$ is real, the sine transform vanishes at the maximum of the global cosine transform.  The sine transform then has a maximum on the right and a minimum on the left.  The opposition of the phases of the Fourier transforms of $P_{13}$ and $P_{23}$ at the 1-2 oscillation maxima again imply that these extrema of the sine transform of the full spectrum roughly coincide with the extrema of the sine transform of $P_{13}$ alone, as can be seen at 58 km in Fig. \ref{sinfig}.  They simply correspond to the values of $k$ for which the phase in (\ref{segno}) is $\pm\pi/2$
\beq
k^L=\frac{\mn31}{2}-\frac{\m21}{4}\hsp
k^R=\frac{\mn31}{2}+\frac{\m21}{4}.
\eeq
Defining the sine transform so as to take the imaginary part of $F$ with the same normalization as for the cosine transform one then finds
\beq
F_{\sin}(k^L)=-i\left(e^{-\beta/4}-\tp212 e^{-\frac{\beta}{4}(2\mp 1)^2}\right)\hsp
F_{\sin}(k^R)=i\left(e^{-\beta/4}-\tp212 e^{-\frac{\beta}{4}(2\pm 1)^2}\right)
\eeq
where the upper sign corresponds to the normal hierarchy.  In the case of the normal hierarchy the second term in the minimum on the left is larger and so $|F_{\sin}(k^L)|<|F_{\sin}(k^R)|$, whereas in the case of the inverted hierarchy the maximum on the right is larger than the minimum on the left.  Thus we have reproduced the correlation between the hierarchy and the relative sizes of these extrema observed in Ref. \cite{caojun}.

Note that in the case of the normal hierarchy both extrema are more positive and in the case of the inverted hierarchy both are more negative, in other words {\textbf{near its maximum $F$ has a positive (negative) imaginary part in the case of a normal (inverted) hierarchy}}.  This provides a new test which allows one to extrapolate the hierarchy from the Fourier transform of the survival probability.  Of course the optimal indicator of the hierarchy will be a weighted sum of these three tests, with weights that may be determined by series of simulations.  The ability to distinguish the hierarchies may also be improved by convoluting the observed spectrum with a slowly varying function of $(L/E)$ which weights neutrinos near the 1-2 maximum more heavily.  One can also use simulated data to determine which weighting functions are optimal for this task.

\subsection{Nonlinear Fourier transform}
The Fourier transform methods described above essentially work because the phase at the maximum is determined by the hierarchy, positive for the normal hierarchy and negative for the inverted hierarchy.  In the case of just $1-3$ oscillations, this phase is zero, since the corresponding oscillations are a cosine and the real part of the Fourier transform is determined by the cosine transform.  However when $P_{23}$ is included the peaks of the untransformed spectrum move according to Eq. (\ref{pichi}).  The distance between the untransformed peaks changes, which in general  moves the Fourier transformed peak.  

Critically, the untransformed $P_{13}+P_{23}$ also loses its mod $2/\mn31$ periodicity, as $\an$ is not a linear function of $n$.  A given detector is only sensitive to some of the peaks, corresponding to a certain region in Fig. \ref{anfig}.  Such regions are generally dominated by a domain in which $\an$ can be approximated not by a linear function, but by a linear function plus a constant offset.  This constant offset implies that the convolution of $P_{13}+P_{23}$ with the reactor neutrino spectrum is not of the form $\cos(\kappa L/E)$ but rather of the form $\cos(\kappa L/E+c)$ where the sign of $c$ is determined by the hierarchy.  This offset, $c$, leads to a translation in $L/E$ space which, after the Fourier transform, becomes a phase.  Thus the hierarchy determines an overall phase of the Fourier transform, which leads to the observable indicators described in the previous subsection.

The Fourier transform method is robust since it sums multiple peaks together, and so it requires less neutrinos than a direct analysis of the positions of the peaks.  However it is inefficient because, as was just described, it works by approximating the function $\an$ to be affine in the energy range which is probed.  So one might ask if the performance would be increased by performing not an ordinary Fourier transform, but a Fourier transform with the nonlinearity of $\an$ built in.  The nonlinearity depends upon $n\pm\an$ which depends on the hierarchy and also weakly upon the neutrino mass matrix.  One may therefore attempt a nonlinear Fourier transform with both choices of hierarchy, and if desired, a weight function $g(L/E)$.  Such a nonlinear cosine transform is given by
\beq
F(k)=\int d\left(\frac{L}{E}\right)\Phi(L/E) P_{ee}(L/E) g(L/E) \cos\left(k\frac{L}{E}\mp 2\pi\alpha\left(\frac{k}{2\pi}\frac{L}{E}\right)\right)
\eeq
where $\alpha(n)$ is a possibly hierarchy-dependent function which interpolates between the discrete values of $\an$ and the positive sign corresponds to the inverted hierarchy, such as
\beq
\alpha(n)=\frac{1}{2\pi}\mathrm{arctan}\left[\frac{\sin(2\pi\epsilon n)}{(1\pm\epsilon){\mathrm{cot}}^2(\th12)+\cos(2\pi\epsilon n)}\right].
\eeq
This transform will add all of the peaks together with the same phase for the correct hierarchy whereas the peaks will be distorted by the other hierarchy.  Therefore the correct hierarchy can be determined from the fact that the corresponding nonlinear Fourier transform will have a larger absolute maximum.

\section{Interference effects} \label{intsez}

\subsection{Interference between reactors separated by 1 km} \label{unoproblema}

Consider two reactors of equal thermal power separated by a distance $D$.  If a detector is located at a distance $L>>D$ from the nearest and if the baseline makes an angle $\theta$ with respect to the line passing through both reactors, then the distance from the detector to the far reactor will be $L+d$ where $d=D\cos(\theta)$. 

Adding the flux from both reactors, one finds that the 1-3 oscillations interfere 
\beq
P_{13}\propto \cos\left(\frac{\mn31L}{2E}\right)+ \cos\left(\frac{\mn31(L+d)}{2E}\right)=2 \cos\left(\frac{\mn31d}{4E}\right)\cos\left(\frac{\mn31(L+d/2)}{2E}\right).
\eeq
The neutrinos from the two sources are not coherent, it is not the wavefunctions that add, but the probabilities.  And the result is that the amplitude of the oscillations at energy $E$ is suppressed by a factor of
\beq
\cos\left(\frac{\mn31d}{4E}\right)= \cos\left(3\frac{d/\mathrm{km}}{E/\mathrm{MeV}}\right).
\eeq
In particular they annihilate entirely when
\beq
\frac{d}{\mathrm{km}}=0.5 \frac{E}{\mathrm{MeV}}.
\eeq

\begin{figure} 
\begin{center}
\includegraphics[width=6in,height=2.7in]{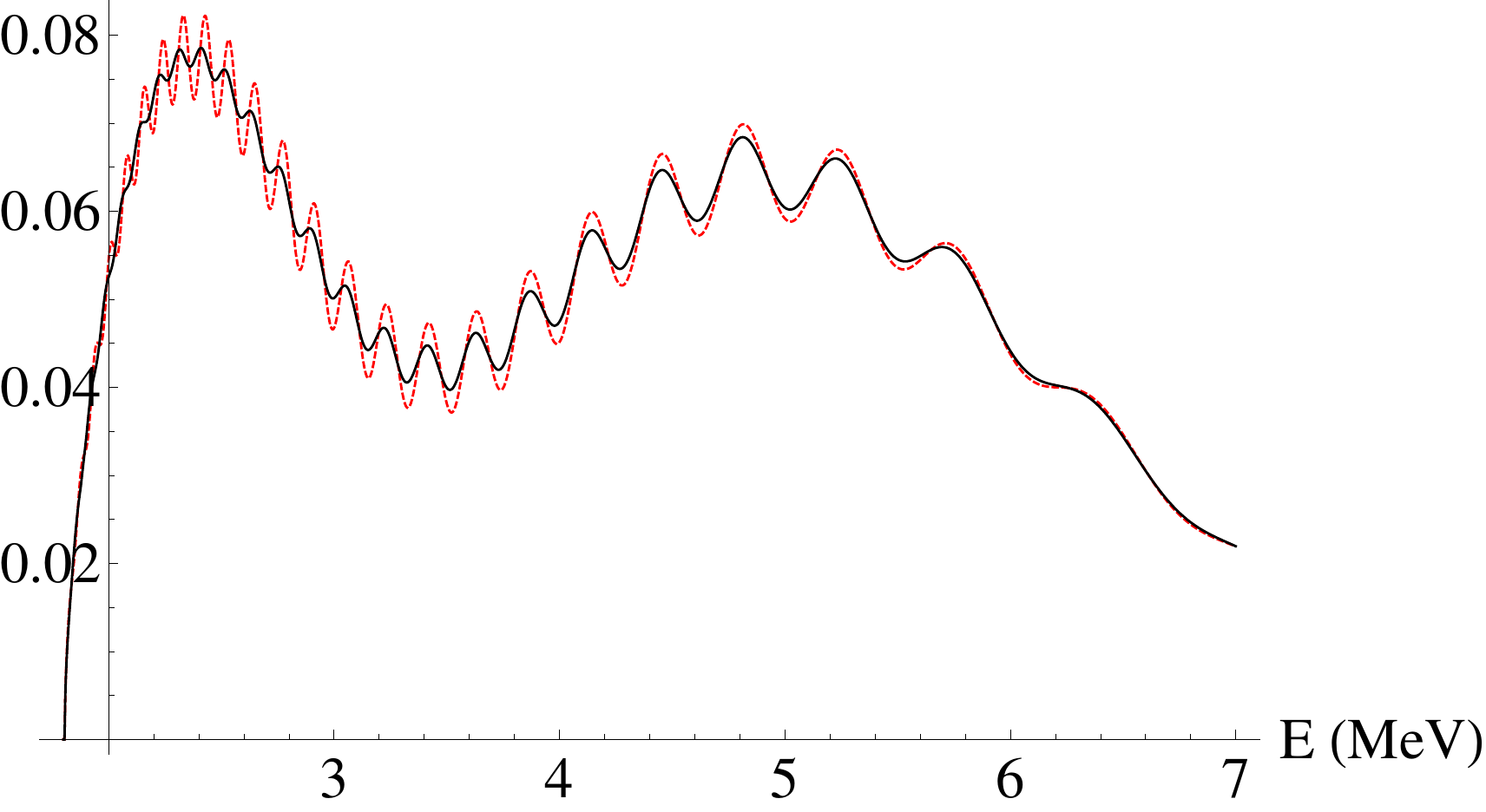}
\caption{The spectrum observed at the Daya Bay location suggested in Ref. \cite{caojunseminario} assuming the normal hierarchy.  The solid black curve includes the interference between Daya Bay and Ling Ao reactors and the red dashed curve does not.  It is assumed that the layout of the Haifeng reactor complex yields as much interference as that at Daya Bay and Ling Ao.}
\label{intfig}
\end{center}
\end{figure}

Consider for example the pair of Daya Bay reactors and the two pairs of Ling Ao reactors which lie along a line at a distance of 0.8 and 1.4 km from the Daya Bay reactors.  The Daya Bay II reactor location suggested in Ref. \cite{caojunseminario} is at an angle of 15 degrees with respect to a continuation of this line, yielding $d=$1.2\ km for the nearest and furthest reactor pair.  The resulting change in the spectrum is shown in Fig. \ref{intfig}.  The 1-3 oscillations at 2.4 MeV completely cancel between these two reactor pairs, leaving only the oscillations of the middle reactor, and so effectively damping the oscillation amplitude by a factor of 3, which implies that an equally precise measurement of a peak at that energy requires 9 times as much flux as it would have without interference.

At the peak energy of 3.6 MeV the annihilation is not complete, but the 1-3 oscillation amplitude of neutrinos coming from the nearest and farthest detectors is reduced by a factor of $\cos(1)=0.6$.  Thus the total amplitude of the peak from all three reactor pairs is reduced by about 30\%, and so 70\% more neutrino flux will be necessary to observe these peaks.

This problem can be avoided if the detector is equidistant from all of the reactors.  In the case of the Daya Bay and Ling Ao complex, in the case of RENO and somewhat trivially in the case of Double Chooz this is possible as all of the reactors essentially lie along a line.  It means however that such a detector will not be equidistant from any reactor that may eventually be built at Haifeng or Huizhou, thus reducing the neutrino fluxes assumed in Refs. \cite{caojun,caojunseminario,yifangseminario} by a factor of 2.  Worse yet, the Haifeng and Huizhou reactors then provide a strong and undesirable background, as will be described in Subsec. \ref{centoproblema}.

\subsection{Interference between reactors separated by 100 km} \label{centoproblema}

The 1-2 oscillation maximum
\beq
L=\frac{2\pi E}{\m21}=17\left(\frac{E}{\mathrm{MeV}}\right)\mathrm{km}
\eeq
provides an ideal baseline to determine the mass hierarchy for a number of reasons.  Among these is that while the $P_{13}+P_{23}$ oscillation amplitude of 
$(\cp212-\sp212)\spp2213$ is a factor of 3 less than its amplitude $\spp2213$ at the 1-2 maxima, the flux remaining after 1-2 oscillations is also smaller by a factor of $1/(1-\sp212\cp413)$ which is about 5.  Thus the 1-3 oscillations are a larger fraction of the total flux, making them easier to see above systematic errors, although not above statistical errors as $3>\sqrt{5}$.

However the smaller signal and smaller flux means that this part of the spectrum is particularly prone to interference from distant reactors, in particular those that are near the first 1-2 oscillation minimum
\beq
L=\frac{4\pi E}{\mn31}=32\left(\frac{E}{\mathrm{MeV}}\right)\mathrm{km}.
\eeq
In this case the additional factor of 5 in flux from the distant reactor outweighs the factor of 4 distance suppression, and so if both reactor complexes have the same strength then most of the flux at the 1-2 maximum will be background.  This means that to obtain the same energy resolution at the 1-2 maximum peaks one will need more than 4 times as much neutrino flux.  

This is the case for example with a detector placed 58 km away from the Daya Bay/Ling Ao complex, perpendicular to the reactors.  It would be at the 1-2 minimum of the proposed Haifeng and Huizhou reactors and would also suffer significant contamination from the Taishan and Yangjiang reactor complexes, at each of which at least 3 reactors are already under construction.  Similarly a reactor placed 60 km from Daya Bay, Ling Ao and Haifeng would be at the 1-2 minimum of the proposed 17.4 GW thermal power reactor complex at Lufeng.

As this effect increases the flux at the 1-2 oscillation maximum, it is also a serious obstacle to an accurate measurement of $\theta_{12}$.  If a model of the background neutrino flux from distant reactors is wrong, the error in the total flux can be compensated for by an error in the determination of $\theta_{12}$.  Ideally this problem can be solved by using two medium baseline detectors instead of one, which would break this degeneracy.  However in a world limited by costs, it may instead be necessary to simply try to make the background from other reactors as small as possible, thus minimizing the potential error in the determination of the hierarchy and in the determination of $\theta_{12}$.

Unlike the short distance interference problem discussed in Subsec. \ref{unoproblema}, the fractional contamination caused by distant reactors depends on the baseline $L$ to the reactor whose neutrinos provide the signal.  The fractional contamination from undesired reactor neutrinos is inversely proportional to the desired signal strength, and so it is proportional to $L^2$.  Thus this problem is minimized by placing the detector as near to the reactor as possible.  If there is only one detector, and one wishes to use it to determine the hierarchy, then as explained in Subsec. \ref{cubicosez}, this minimum distance cannot be shorter than about 45 km.

\section{Conclusions}

The newly discovered high value of $\theta_{13}$ means that the determination of the neutrino mass hierarchy at a medium baseline reactor experiment is now practical.  Previous analyses of such experiments have assumed values of $\spp2213$ an order of magnitude or more below its true value.  At such low values, individual peaks of the electron antineutrino spectra could not be resolved and one instead needed to rely upon a Fourier analysis \cite{hawaii,caojun}.  

In this note we have reconsidered the determination of the neutrino mass hierarchy now that 1-3 oscillations are large and so individual 1-3 peaks in the spectrum can easily be observed.  We found that the position of each peak determines a particular combination of the neutrino mass differences, for example the first 10 peaks all determine the same difference $\cp212\mn31+\sp212\mn32$.  In particular this means that the first 10 peaks alone cannot be used to determine the hierarchy, as they are fit equally well by the wrong hierarchy model in which this mass difference is preserved.  On the other hand we found that the positions of the next 10 peaks depend on distinct combinations of the mass differences, and so combining the energies at different peaks with some beyond the $10$th can lead to a determination of the hierarchy.  We also estimated the detector resolution and neutrino flux which are needed for such a goal, although an accurate determination will be left for the simulations to be discussed in our companion paper.

The information about the hierarchy is therefore contained in the low energy peaks, which suffer the most from the effects of poor resolution,  low neutrino flux and interference.  While the individual peaks are somewhat difficult to resolve, the situation can nonetheless be improved with a Fourier transform.  We have analyzed the Fourier transform of the spectrum, deriving the phenomenological hierarchy indicators proposed in Refs. \cite{hawaii,caojun} and providing a new indicator, the complex phase at the peak of the Fourier transform, which we claim will be positive for the normal hierarchy and negative for the inverse hierarchy.  We also propose a new hierarchy-dependent nonlinear Fourier transform which will lead to a higher peak using the transform corresponding to the correct hierarchy.  Our simulations show that an analysis which combines the linear and nonlinear transform methods is able to identify the hierarchy considerably more reliably than one which relies entirely upon the linear transform.

The strength of the Fourier transform method is that it sums together all of the peaks to increase the strength of the signal with respect to the noise.  If the energy spectrum is shifted, this simply leads to an overall phase in the Fourier transform which does not seriously affect the analysis.  Likewise a uniform stretching of the spectrum simply shifts the peaks, which does not affect the determination of the hierarchy at all.  Any nonlinearity however poses a much more serious problem, as it can lead to an interference between the peaks in the spectrum which mutates or destroys the peak structure of the Fourier transform.  If the nature of the nonlinearity is known then one can adjust the analysis to correct for it \cite{petcov2010} however if the nonlinearity is known it could be corrected directly in the reading of the energy.  In general the nonlinearity of the response is only known at energies where the detector has been calibrated with radioactive sources.  It may therefore be desirable to modify the Fourier transform so as to weigh the more reliable energies more heavily.  One may also wish to weigh the energies near the 1-2 maximum more heavily, as they contribute few neutrinos but are indispensable in a determination of the hierarchy.  These  weights can be optimized by testing various hierarchy determination algorithms against simulated data.

We also discussed two interference effects.  First of all, reactors in the same array are generally separated by distances of order 1 km, which means that neutrinos arriving at a detector from one reactor at a 1-3 maximum may be at the same energy as those arriving from another at a 1-3 minimum.  The result is that the 1-3 oscillation signal can be severely reduced at the low energies in which this oscillation can occur within a reactor complex.  These are just the low energies which are necessary for the determination of the hierarchy, and so this interference poses a serious problem.  One solution is to place the detectors orthogonal to arrays of reactors.

The second interference effect arises from the fact that while the 1-2 oscillation maximum is the most useful energy range at which to determine the hierarchy, it enjoys a much lower neutrino flux than the 1-2 oscillation minimum.  As a result at these energy ranges one can expect serious contamination from distant reactors at their 1-2 minimum.  This problem, as well as the degeneracy between the neutrino flux from distant reactors and $\theta_{12}$, is optimally solved by using two detectors at different baselines.  However a cheaper solution to the same problem is to use a single detector at a shorter baseline, such as 45 km NNW of Daya Bay under  B\'aiY\'unZh\`ang or, 10 km further west, under the mountains 52 km from Daya Bay.


\section* {Acknowledgement}

\noindent
JE is supported by the Chinese Academy of Sciences
Fellowship for Young International Scientists grant number
2010Y2JA01. EC and XZ are supported in part by the NSF of
China.  


\end{document}

\bibitem{lsnd}
A.~Aguilar-Arevalo {\it et al.}  [LSND Collaboration],
  ``Evidence for neutrino oscillations from the observation of anti-neutrino(electron) appearance in a anti-neutrino(muon) beam,''
  Phys.\ Rev.\ D {\bf 64} (2001) 112007
  [hep-ex/0104049].

\bibitem{minibooneanom}
A.~A.~Aguilar-Arevalo {\it et al.}  [The MiniBooNE Collaboration],
  ``A Search for electron neutrino appearance at the $\Delta m^{2} \sim 1$eV$^{2}$ scale,''
  Phys.\ Rev.\ Lett.\  {\bf 98} (2007) 231801
  [arXiv:0704.1500 [hep-ex]].
A.~A.~Aguilar-Arevalo {\it et al.}  [MiniBooNE Collaboration],
  ``Unexplained Excess of Electron-Like Events From a 1-GeV Neutrino Beam,''
  Phys.\ Rev.\ Lett.\  {\bf 102} (2009) 101802
  [arXiv:0812.2243 [hep-ex]].
 A.~A.~Aguilar-Arevalo {\it et al.}  [The MiniBooNE Collaboration],
  ``Event Excess in the MiniBooNE Search for $\bar \nu_\mu \rightarrow \bar \nu_e$ Oscillations,''
  Phys.\ Rev.\ Lett.\  {\bf 105} (2010) 181801
  [arXiv:1007.1150 [hep-ex]].

\bibitem{minosanom}
P.~Adamson {\it et al.}  [MINOS Collaboration],
  ``First direct observation of muon antineutrino disappearance,''
  Phys.\ Rev.\ Lett.\  {\bf 107} (2011) 021801
  [arXiv:1104.0344 [hep-ex]].

\bibitem{zichichi}
A.~Zichichi,
``Results from LVD-OPERA Combined Analysis: A Time-Shift in the OPERA Setup,"
available online at http://agenda.infn.it/getFile.py/access?resId=0\&materialId=slides\&confId=4896.

\bibitem{miniboonetuttobene}
E.~D.~Zimmerman [MiniBooNE Collaboration],
  ``Updated Search for Electron Antineutrino Appearance at MiniBooNE,''
  arXiv:1111.1375 [hep-ex].

\bibitem{minostuttobene}
P.~Adamson {\it et al.}  [MINOS Collaboration],
  ``An improved measurement of muon antineutrino disappearance in MINOS,''
  arXiv:1202.2772 [hep-ex].

\bibitem{nuovoflusso}
T.~.A.~Mueller, D.~Lhuillier, M.~Fallot, A.~Letourneau, S.~Cormon, M.~Fechner, L.~Giot and T.~Lasserre {\it et al.},
  ``Improved Predictions of Reactor Antineutrino Spectra,''
  Phys.\ Rev.\ C {\bf 83} (2011) 054615
  [arXiv:1101.2663 [hep-ex]].
P.~Huber,
  ``On the determination of anti-neutrino spectra from nuclear reactors,''
  Phys.\ Rev.\ C {\bf 84} (2011) 024617
   [Erratum-ibid.\ C {\bf 85} (2012) 029901]
  [arXiv:1106.0687 [hep-ph]].

\bibitem{reattoreanom}
G.~Mention, M.~Fechner, T. Lasserre, T~.A.~Mueller, D.~Lhuillier, M.~Cribier and A.~Letourneau,
  ``The Reactor Antineutrino Anomaly,''
  Phys.\ Rev.\ D {\bf 83} (2011) 073006
  [arXiv:1101.2755 [hep-ex]].

\bibitem{smirnov}
P.~C.~de Holanda and A.~Y.~.Smirnov,
  ``Homestake result, sterile neutrinos and low-energy solar neutrino experiments,''
  Phys.\ Rev.\ D {\bf 69} (2004) 113002
  [hep-ph/0307266].
P.~C.~de Holanda and A.~Y.~.Smirnov,
  ``Solar neutrino spectrum, sterile neutrinos and additional radiation in the Universe,''
  Phys.\ Rev.\ D {\bf 83} (2011) 113011
  [arXiv:1012.5627 [hep-ph]].

\bibitem{icecube}
R.~Abbasi {\it et al.}  [IceCube Collaboration],
  ``Measurement of the atmospheric neutrino energy spectrum from 100 GeV to 400 TeV with IceCube,''
  Phys.\ Rev.\ D {\bf 83} (2011) 012001
  [arXiv:1010.3980 [astro-ph.HE]].

\bibitem{giuntireview}
  C.~Giunti and M.~Laveder,
  ``Implications of 3+1 Short-Baseline Neutrino Oscillations,''
  Phys.\ Lett.\ B {\bf 706} (2011) 200
  [arXiv:1111.1069 [hep-ph]].

\bibitem{sterilecosm}
J.~Hamann, S.~Hannestad, G.~G.~Raffelt and Y.~Y.~Y.~Wong,
  ``Sterile neutrinos with eV masses in cosmology: How disfavoured exactly?,''
  JCAP {\bf 1109} (2011) 034
  [arXiv:1108.4136 [astro-ph.CO]].

\bibitem{dayabay}
  F.~P.~An {\it et al.}  [DAYA-BAY Collaboration],
  ``Observation of electron-antineutrino disappearance at Daya Bay,''
  Phys.\ Rev.\ Lett.\  {\bf 108} (2012) 171803
  [arXiv:1203.1669 [hep-ex]].

\bibitem{piureattori}
L.~A.~Mikaelyan and V.~V.~Sinev,
  ``Neutrino oscillations at reactors: What next?,''
  Phys.\ Atom.\ Nucl.\  {\bf 63} (2000) 1002
   [Yad.\ Fiz.\  {\bf 63N6} (2000) 1077]
  [hep-ex/9908047].

\bibitem{neut2012}
D. Dwyer,
``Daya Bay Results," presented at Neutrino 2012 in Kyoto.
Available at http://neu2012.kek.jp/neu2012/programme.html.

\bibitem{doublechooz}
Y.~Abe {\it et al.}  [DOUBLE-Chooz Collaboration],
  ``Indication for the disappearance of reactor electron antineutrinos in the Double Chooz experiment,''
  Phys.\ Rev.\ Lett.\  {\bf 108} (2012) 131801
  [arXiv:1112.6353 [hep-ex]].

\bibitem{reno}
  J.~K.~Ahn {\it et al.}  [RENO Collaboration],
  ``Observation of Reactor Electron Antineutrino Disappearance in the RENO Experiment,''
  Phys.\ Rev.\ Lett.\  {\bf 108} (2012) 191802
  [arXiv:1204.0626 [hep-ex]].

\bibitem{nuturn}
``Observation of reactor neutrino disappearance at RENO," presented at $\nu$TURN 2012 under Gran Sasso.  Available at http://agenda.infn.it/contributionListDisplay.py?confId=4722.

\bibitem{globale1}
G.~L.~Fogli, E.~Lisi, A.~Marrone, A.~Palazzo and A.~M.~Rotunno,
  ``Evidence of $\theta_{13}$>0 from global neutrino data analysis,''
  Phys.\ Rev.\ D {\bf 84} (2011) 053007
  [arXiv:1106.6028 [hep-ph]].

\bibitem{globale2}
  T.~Schwetz, M.~Tortola and J.~W.~F.~Valle,
  ``Where we are on $\theta_{13}$: addendum to 'Global neutrino data and recent reactor fluxes: status of three-flavour oscillation parameters',''
  New J.\ Phys.\  {\bf 13} (2011) 109401
  [arXiv:1108.1376 [hep-ph]].

\bibitem{paloverde}
F.~Boehm, J.~Busenitz, B.~Cook, G.~Gratta, H.~Henrikson, J.~Kornis, D.~Lawrence and K.~B.~Lee {\it et al.},
  ``Final results from the Palo Verde neutrino oscillation experiment,''
  Phys.\ Rev.\ D {\bf 64} (2001) 112001
  [hep-ex/0107009].

\bibitem{chooz}
M.~Apollonio {\it et al.}  [Chooz Collaboration],
  ``Search for neutrino oscillations on a long baseline at the Chooz nuclear power station,''
  Eur.\ Phys.\ J.\ C {\bf 27} (2003) 331
  [hep-ex/0301017].

\bibitem{neutdarke}
X.~-J.~Bi, P.~-H.~Gu, X.~-l.~Wang and X.~-M.~Zhang,
  ``Thermal leptogenesis in a model with mass varying neutrinos,''
  Phys.\ Rev.\ D {\bf 69} (2004) 113007
  [hep-ph/0311022].
  R.~Takahashi and M.~Tanimoto,
  ``Model of mass varying neutrinos in SUSY,''
  Phys.\ Lett.\ B {\bf 633} (2006) 675
  [hep-ph/0507142].
  R.~Takahashi and M.~Tanimoto,
  ``Speed of sound in the mass varying neutrinos scenario,''
  JHEP {\bf 0605} (2006) 021
  [astro-ph/0601119].
  E.~Ciuffoli, J.~Evslin, J.~Liu and X.~Zhang,
  ``OPERA and a Neutrino Dark Energy Model,''
  arXiv:1109.6641 [hep-ph].

\bibitem{neal04}
  D.~B.~Kaplan, A.~E.~Nelson and N.~Weiner,
  ``Neutrino oscillations as a probe of dark energy,''
  Phys.\ Rev.\ Lett.\  {\bf 93} (2004) 091801
  [hep-ph/0401099].

\bibitem{tortola}
  M.~Tortola, J.~W.~F.~Valle and D.~Vanegas,
  ``Global status of neutrino oscillation parameters after recent reactor measurements,''
  arXiv:1205.4018 [hep-ph].

\bibitem{foglinuovo}
G.L. Fogli, E. Lisi, A. Marrone, D. Montanino, A. Palazzo and A.M. Rotunno,
``Global analysis of neutrino masses, mixings and phases: entering the era of leptonic CP violation searches,"
arXiv:1205.5254 [hep-ph].

\bibitem{bugey4}
Y.~Declais, H.~de Kerret, B.~Lefievre, M.~Obolensky, A.~Etenko, Y.~.Kozlov, I.~Machulin and V.~Martemyanov {\it et al.},
  ``Study of reactor anti-neutrino interaction with proton at Bugey nuclear power plant,''
  Phys.\ Lett.\ B {\bf 338} (1994) 383.

\bibitem{dayafeb}
F.~P.~An {\it et al.}  [Daya Bay Collaboration],
  ``A side-by-side comparison of Daya Bay antineutrino detectors,''
  arXiv:1202.6181 [physics.ins-det].

\bibitem{daya2007}
  X.~Guo {\it et al.}  [Daya-Bay Collaboration],
  ``A Precision measurement of the neutrino mixing angle theta(13) using reactor antineutrinos at Daya-Bay,''
  hep-ex/0701029.


\bibitem{tredueterm}
 A.~Melchiorri, O.~Mena, S.~Palomares-Ruiz, S.~Pascoli, A.~Slosar and M.~Sorel,
  ``Sterile Neutrinos in Light of Recent Cosmological and Oscillation Data: A Multi-Flavor Scheme Approach,''
  JCAP {\bf 0901} (2009) 036
  [arXiv:0810.5133 [hep-ph]].

\bibitem{neutrinoasym}
  S.~Hannestad, I.~Tamborra and T.~Tram,
  ``Thermalisation of light sterile neutrinos in the early universe,''
  arXiv:1204.5861 [astro-ph.CO].

\bibitem{baoscoperta}
  D.~J.~Eisenstein {\it et al.}  [SDSS Collaboration],
  ``Detection of the baryon acoustic peak in the large-scale correlation function of SDSS luminous red galaxies,''
  Astrophys.\ J.\  {\bf 633} (2005) 560
  [astro-ph/0501171].


\bibitem{bao}
W.~J.~Percival, S.~Cole, D.~J.~Eisenstein, R.~C.~Nichol, J.~A.~Peacock, A.~C.~Pope and A.~S.~Szalay,
  ``Measuring the Baryon Acoustic Oscillation scale using the SDSS and 2dFGRS,''
  Mon.\ Not.\ Roy.\ Astron.\ Soc.\  {\bf 381} (2007) 1053
  [arXiv:0705.3323 [astro-ph]].

\bibitem{cosmomc}
  A.~Lewis and S.~Bridle,
  ``Cosmological parameters from CMB and other data: A Monte Carlo approach,''
  Phys.\ Rev.\ D {\bf 66} (2002) 103511
  [astro-ph/0205436].

\bibitem{wmap}
E.~Komatsu {\it et al.}  [WMAP Collaboration],
  ``Seven-Year Wilkinson Microwave Anisotropy Probe (WMAP) Observations: Cosmological Interpretation,''
  Astrophys.\ J.\ Suppl.\  {\bf 192} (2011) 18
  [arXiv:1001.4538 [astro-ph.CO]].

\bibitem{Suzuki:2011hu}
  N.~Suzuki {\it et al.},
  ``The Hubble Space Telescope Cluster Supernova Survey: V. Improving the Dark
  Energy Constraints Above z>1 and Building an Early-Type-Hosted Supernova
  Sample,''
  arXiv:1105.3470 [astro-ph.CO].

\bibitem{h}
A.~G.~Riess, L.~Macri, S.~Casertano, H.~Lampeitl, H.~C.~Ferguson, A.~V.~Filippenko, S.~W.~Jha and W.~Li {\it et al.},
  ``A 3\% Solution: Determination of the Hubble Constant with the Hubble Space Telescope and Wide Field Camera 3,''
  Astrophys.\ J.\  {\bf 730} (2011) 119
   [Erratum-ibid.\  {\bf 732} (2011) 129]
  [arXiv:1103.2976 [astro-ph.CO]].

\bibitem{nessundivergenza}
J.~-Q.~Xia, G.~-B.~Zhao, B.~Feng, H.~Li and X.~Zhang,
  ``Observing dark energy dynamics with supernova, microwave background and galaxy clustering,''
  Phys.\ Rev.\ D {\bf 73} (2006) 063521
  [astro-ph/0511625].
G.~-B.~Zhao, J.~-Q.~Xia, M.~Li, B.~Feng and X.~Zhang,
  ``Perturbations of the quintom models of dark energy and the effects on observations,''
  Phys.\ Rev.\ D {\bf 72} (2005) 123515
  [astro-ph/0507482].

\bibitem{altroquintom}
  E.~Giusarma, M.~Archidiacono, R.~de Putter, A.~Melchiorri and O.~Mena,
  ``Sterile neutrino models and nonminimal cosmologies,''
  Phys.\ Rev.\ D {\bf 85} (2012) 083522
  [arXiv:1112.4661 [astro-ph.CO]].

\bibitem{unione2}
N.~Suzuki, D.~Rubin, C.~Lidman, G.~Aldering, R.~Amanullah, K.~Barbary, L.~F.~Barrientos and J.~Botyanszki {\it et al.},
  ``The Hubble Space Telescope Cluster Supernova Survey: V. Improving the Dark Energy Constraints Above z=1 and Building an Early-Type-Hosted Supernova Sample,''
  Astrophys.\ J.\  {\bf 746} (2012) 85
  [arXiv:1105.3470 [astro-ph.CO]].

\bibitem{quintom}
B.~Feng, M.~Li, Y.~-S.~Piao and X.~Zhang,
  ``Oscillating quintom and the recurrent universe,''
  Phys.\ Lett.\ B {\bf 634} (2006) 101
  [astro-ph/0407432].
  B.~Feng, X.~L.~Wang and X.~M.~Zhang, 
``Dark energy constraints from the cosmic age and supernova,''
 Phys.\ Lett.\  B {\bf 607} (2005) 35  
 [arXiv:astro-ph/0404224].
  X.~-F.~Zhang, H.~Li, Y.~-S.~Piao and X.~-M.~Zhang,
``Two-field models of dark energy with equation of state across -1,''
  Mod.\ Phys.\ Lett.\ A {\bf 21}, 231 (2006)
  [astro-ph/0501652].

\bibitem{neutrinoasymvecchio}
  R.~Foot and R.~R.~Volkas,
  ``Reconciling sterile neutrinos with big bang nucleosynthesis,''
  Phys.\ Rev.\ Lett.\  {\bf 75} (1995) 4350
  [hep-ph/9508275].

\bibitem{fr}
  H.~Motohashi, A.~A.~Starobinsky and J.~'i.~Yokoyama,
  ``Cosmology based on f(R) Gravity admits 1 eV Sterile Neutrinos,''
  arXiv:1203.6828 [astro-ph.CO].

\bibitem{robert}
R.~H.~Brandenberger, N.~Kaiser, D.~N.~Schramm and N.~Turok,
  ``Galaxy and Structure Formation with Hot Dark Matter and Cosmic Strings,''
  Phys.\ Rev.\ Lett.\  {\bf 59} (1987) 2371.
R.~H.~Brandenberger, A.~Mazumdar and M.~Yamaguchi,
  ``A Note on the robustness of the neutrino mass bounds from cosmology,''
  Phys.\ Rev.\ D {\bf 69} (2004) 081301
  [hep-ph/0401239].

\bibitem{monopoli}
J.~Evslin and S.~B.~Gudnason,
  ``High Q BPS Monopole Bags are Urchins,''
  arXiv:1111.3891 [hep-th].
J.~Evslin and S.~B.~Gudnason,
  ``Dwarf Galaxy Sized Monopoles as Dark Matter?,''
  arXiv:1202.0560 [astro-ph.CO].

\bibitem{lsndnonstandard}
  G.~Karagiorgi, M.~H.~Shaevitz and J.~M.~Conrad,
  ``Confronting the short-baseline oscillation anomalies with a single sterile neutrino and non-standard matter effects,''
  arXiv:1202.1024 [hep-ph].

\bibitem{envir}
  G.~Karagiorgi, M.~H.~Shaevitz and J.~M.~Conrad,
  ``Confronting the short-baseline oscillation anomalies with a single sterile neutrino and non-standard matter effects,''
  arXiv:1202.1024 [hep-ph].

\end{thebibliography}

\end{document}